\documentclass[preprint,12pt]{elsarticle}

\usepackage{graphicx}
\usepackage{amssymb}
\usepackage[fleqn]{amsmath}
\usepackage{algorithm}
\usepackage{algpseudocode}

\usepackage{subcaption}

\usepackage{lineno}

\journal{  }

\begin{document}

\begin{frontmatter}

\title{On the performance of a highly-scalable Computational Fluid Dynamics code on AMD, ARM and Intel processors}
% Performance and scalability comparative of a CFD code on ARM and x86 technologies
%Performance and scalability of ARM, AMD and Intel processors for Computational Fluid Dynamics 

\author[1,2]{Pablo Ouro\corref{cor1}} \ead{pablo.ouro@manchester.ac.uk}
\address[1]{School of Mechanical, Aerospace and Civil Engineering, University of Manchester, Manchester, M13 9PL, UK}
\address[2]{Hydro-environmental Research Centre, School of Engineering, Cardiff University, CF24 3AA Cardiff, UK}
\cortext[cor1]{Corresponding author: Dr Pablo Ouro (pablo.ouro@manchester.ac.uk)}
\author[3]{Unai Lopez-Novoa} \ead{unai.lopez@ehu.eus}
\address[3]{Department of Computer Languages and Systems, University of the Basque Country, Spain}
\author[4]{Martyn F. Guest} \ead{GuestMF@cardiff.ac.uk}
\address[4]{Advanced Research Computing, Cardiff University, Cardiff, UK}
%\author[4]{Simon McIntosh-Smith} \ead{s.mcintosh-smith@bristol.ac.uk}
%\address[4]{High Performance Computing research group, Department of Computer Science, University of Bristol, Bristol, UK}

\begin{abstract}

No area of computing is hungrier for performance than High Performance Computing (HPC), the demands of which continue to be a major driver for processor performance and adoption of accelerators, and also advances in memory, storage, and networking technologies. A key feature of the Intel processor domination of the past decade has been the extensive adoption of GPUs as coprocessors, whilst more recent developments have seen the increased availability of a number of CPU processors, including the novel ARM-based chips.
This paper analyses the performance and scalability of a state-of-the-art Computational Fluid Dynamics (CFD) code on three HPC cluster systems equipped with AMD EPYC-Rome (EPYC, 4096 cores), ARM-based Marvell ThunderX2 (TX2, 8192 cores) and Intel Skylake (SKL, 8000 cores) processors.
Three benchmark cases are designed with increasing computation-to-communication ratio and numerical complexity, namely lid-driven cavity flow, Taylor-Green vortex and a travelling solitary wave using the level-set method, adopted with $4^{th}$-order central-differences or a $5^{th}$-order WENO scheme.
Our results show that the EPYC cluster delivers the best code performance for all the setups under consideration. In the first two benchmarks, the SKL cluster demonstrates faster computing times than the TX2 system, whilst in the solitary wave simulations, the TX2 cluster achieves good scalability and similar performance to the EPYC system, both improving on that obtained with the SKL cluster. These results suggest that while the Intel SKL cores deliver the best strong scalability, the associated cluster performance is lower compared to the EPYC system.
The TX2 cluster performance is promising considering its recent addition to the HPC portfolio.
%Overall, these findings reflect the state-of-the-art in HPC hardware for CFD and highlight the need for application-specific tuning in HPC codes to maximise the delivered performance from new chips
\end{abstract}
\begin{keyword}
HPC \sep CFD \sep ARM ThunderX2 \sep AMD EPYC-Rome \sep Intel Skylake \sep Performance Evaluation \sep Massively parallel implementation
\end{keyword}
\end{frontmatter}

%\linenumbers

\section{Introduction}  \label{S:1}

The ability to perform large-scale simulations of turbulent flows has often been restricted by the available capacity of High-Performance Computing (HPC) facilities and the inherent limitations in the performance of the associated processors. 
HPC-driven turbulence research has enabled new insights into fundamental theory \citep{Jimenez2020}, and enabled engineers to build digital environments as virtual representations of physical processes in fields such as hydraulics, environmental turbulent flows, or offshore renewable energy \citep{Stoesser2014,Rodi2017,Sotiropoulos2019}. 

For almost two decades, processor clock frequencies have been increasing at very modest rates, favouring the adoption of multi-core CPUs with increasingly complex memory hierarchies. 
Even if this approach has delivered double, four, eight or more times the attainable Flop/s in a chip, rarely is this accompanied by a similar improvement in memory bandwidth, in practice the main bottleneck for Computational Fluid Dynamics (CFD) codes given their memory bound characteristics. 
This was already foreseen in 2003, when the rate of increase in CPU performance still followed Moore's law, by \citet{Jimenez2003} who pointed out that for CFD, memory bandwidth is as limiting a resource as computing power. 
Following the widespread adoption of GPUs as compute platforms over the past decade, an increasing variety of chip architectures are now available for developers to improve the performance of massively parallel codes \cite{6748067}. However, improvements in code performance will of course depend on the end-user application, e.g. a classical molecular dynamics code may be compute-bound \citep{Guest2019} whilst CFD codes are typically memory-bound \cite{Celani2007}.
This calls for heterogeneous HPC systems that comprise a variety of chip architectures, allowing users to maximise the performance of their specific computing applications \citet{Gray2012}.

A new player has recently emerged in the HPC arena: ARM-based CPUs. The latter have dominated the embedded and mobile market due to its low-power oriented design, but manufacturers have now presented desktop and server level CPUs based on this processor architecture \cite{mcintosh2019performance}. Examples include Marvell with its Thunder series of processors and Fujitsu with the A64FX.

The Mont-Blanc project, led by the Barcelona Supercomputing Centre (BSC), pioneered the deployment of ARM chips in HPC environments with a focus on studying energy efficiency. 
The first prototype in 2014 \cite{Rajovic2014}, Tibidabo, comprised 128 nodes with dual-core ARM Cortex-A9 CPU, succeeded in 2018 by the 960 node system with ARM Cortex-A15 CPUs \cite{Oyarzun2018}. 
An increasing number of HPC facilities are now building supercomputers with ARM-based nodes. The most notable is Fugaku\footnote{Fugaku Supercomputer - https://www.r-ccs.riken.jp/en/fugaku}, a supercomputer hosted in the RIKEN Center for Computational Science, Japan, which hosts 7.3 million ARM-based Fujitsu A64FX cores. In the most recent Top500 list at the time of writing, June 2020\footnote{Top500 list - https://www.top500.org}, Fugaku entered at \#1 by running Linpack with 415.5 PFlop/s of performance and operating at 28.3 MW. In the HPCG list of the same month\footnote{HPCG list - https://www.top500.org/lists/hpcg}, Fugaku also entered at \#1 by running the HPCG benchmark at 13.3 PFlop/s.
This outstanding performance evidences the potential of ARM chips for HPC.

The widespread increase in available computational resources has enabled major, well documented advancements in the application of CFD to many engineering fields, notably in hydraulics, atmospheric or environmental turbulent flows.
In 2015, \citet{Sotiropoulos2015} reviewed how supercomputers enabled progress at the frontiers of hydraulics, including problems related to sediment transport, free-surface flows, or renewable energy, all benefiting from having high-resolution grids that appropriately resolve the turbulent scales using Direct-Numerical Simulation (DNS) or Large-Eddy Simulation (LES) \cite{Stoesser2014,Sotiropoulos2019}. 

At present, DNS is restricted to relatively moderate Reynolds numbers (Re) for classical homogeneous isotropic turbulence or wall-bounded channel flows \citep{Jimenez2020}, as the computing demand depends on the required number of degrees of freedom, in the order of Re$^{11/4}_{\lambda}$ \cite{Sagaut}. 
However, its use for engineering applications remains impractical given the inhomogeneous boundary conditions and high Reynolds numbers.
For instance, \citet{Mazzuoli2019} performed a DNS study with over $2\times 10^9$ grid cells and 250,000 spherical particles at a cost of $10^7$ CPU hours, running for approx. 480 days on 64 Intel Ivy Bridge nodes. This makes it unfeasible to increase the flow regime or at least it requires excessive computational power \cite{Mazzuoli2019}.

Alternatively, LES enables the resolution of complex multi-phase flows or simulations in large numerical domains at a moderate-to-high computational cost.
%as a result of the advancements in HPC, which have made computationally intensive high-fidelity CFD simulations doable using a reasonable amount of HPC resources.
Indeed, one of the most complex challenges in hydraulics and environmental turbulent flows, the multi-phase flow simulation using interface-capturing schemes, is amenable to study by LES \cite{Kang2012a,Kara2015,Xie2015}.
In free-surface open-channel flows, adopting the Level-Set Method (LSM) to represent the air-water interface can provide more accurate results than using the classic frictionless rigid-lid boundary condition, but the former comes at five times higher computational expense (\citet{Krosronejad2019}).

%%%%%%%%%%%%%%
This paper studies the performance and scalability of an in-house CFD code on three clusters featuring distinctive node architectures, namely dual-processor AMD EPYC-Rome, ARMv8.1 ThunderX2, and Intel Skylake nodes, denoted hereafter as EPYC, TX2 and SKL respectively. These have distinctive core-per-node counts, underlying architecture and memory bandwidth, each impacting on the performance of CFD codes.
Initially, two well-known numerical benchmarks are adopted for the computation of the incompressible Navier-Stokes equation: the lid-driven cavity flow and Taylor-Green vortex. 
These tests are designed to investigate code performance when using 4$^{th}$-order central differences or 5$^{th}$-order WENO schemes, in application to problems of increasing sizes ranging from four million to a billion grid cells, thereby enabling performance analysis when increasing the computing workload and communication-to-computation ratio.  
A third case involves the propagation of a solitary wave using the level-set method as representative of a complex multi-phase flow engineering application. 

The structure of the paper is as follows: Section \ref{S:2} introduces the in-house code $\mathtt{Hydro3D}$, numerical schemes to compute the fluxes, and the level-set method. The cluster systems are described in Section \ref{S:3}, while the test cases used in the evaluation are outlined in Section \ref{S:4}. Performance and scalability results are presented and discussed in Section \ref{S:results}, with the main conclusions drawn in Section \ref{S:conclusions}.
%the discussion is presented in Section \ref{S:discussion} providing an overview of the obtained results and implications to the CFD community, and 

%%%%%%%%%%%%%%%%%%%%%%%%%%%%%%%%%%%%%%%%%%%%%%%%%
%\vspace{5cm}
%\pagebreak
%\newpage
\section{Hydro3D: an open-source CFD code} \label{S:2}

The software used in this research is $\mathtt{Hydro3D}$ \cite{Cevheri2016,Ouro2019CAF}, an in-house open-source code \cite{Hydro3Dv60} for incompressible, viscous turbulent flows written in $\mathtt{FORTRAN}$ and fully parallelised with MPI. 
$\mathtt{Hydro3D}$ allows the simulation of moving bodies using the immersed boundary method \cite{Kara2015a,Ouro2017CAF,Ouro2019PRF,Ouro2019JFS,Liu2017}, multi-phase flows with the Eulerian-Eulerian approach of the level-set method \cite{Kara2015,McSherry2018,Chua2019} and Eulerian-Lagrangian framework of Lagrangian particle tracking \cite{Fraga2016}, and a local-mesh refinement method \cite{Cevheri2016}. 
A hybrid MPI/OpenMP version to compute the Immersed Boundary Method (IBM) module (\citet{Ouro2019CAF}) showed that the MPI/OpenMP scheme can outperform pure MPI computations when using IBM kernels with large stencils. 
An efficient multi-grid solver is used to compute the Poisson pressure equation, whose details are presented in \citet{Cevheri2016}.

$\mathtt{Hydro3D}$ is based on the Large-Eddy Simulation (LES) approach in which the most energetic and largest flow structures are explicitly resolved, whilst a sub-grid scale method is responsible for modelling the smallest flow scales considered as those smaller than the filter size, in this case the grid resolution $\Delta x_i$ \citep{Stoesser2014}. The governing equations are the spatially filtered Navier-Stokes equations for incompressible viscous flows, that read

\begin{align}
& \frac{\partial \tilde{u}_i}{\partial x_i} = 0 \\
& \frac{\partial \tilde{u}_i}{\partial t} = -\frac{1}{\rho} \frac{\partial p}{\partial x_i} 
- \tilde{u}_j \frac{\partial \tilde{u}_i}{\partial x_j} 
+ (\nu + \nu_t) \frac{\partial^2 \tilde{u}_i}{\partial x_j^2}  + S_i \label{eq:ns2}
\end{align}

Here $\tilde{u}_i$ = $(u,v,w)^T$ is the spatially-filtered velocity vector (for convenience the $(\tilde{\cdot})$ symbol is omitted hereafter), the coordinates vector is $x_i$ = $(x,y,z)^T$, $\rho$ denotes the fluid density, $p$ is the relative pressure, $\nu$ is the kinematic viscosity of the fluid, and $S_i$ is a source term. 
The eddy-viscosity $\nu_t$ is calculated using the Wall-Adapting Local Eddy-viscosity (WALE) sub-grid scale model from \citet{Nicoud} as,

\begin{align}
    \nu_t = ( C_w \Delta x_i)^2 \frac{ \left( S^d_{ij} S^d_{ij} \right)^{3/2} }{
    \left( \overline{S}_{ij} \overline{S}_{ij} \right)^{5/2}  + \left( S^d_{ij} S^d_{ij} \right)^{5/4} 
    }
\end{align}
with $C_w$ being a constant equal to 0.46, $S^d_{ij}$ is the traceless symmetric part of the square of the velocity gradient tensor, and $\overline{S}_{ij} = 0.5 (\partial_j u_i + \partial_i u_j)$ is the strain-rate tensor of velocities. 

The strategy for domain decomposition and MPI communication are described below in Section \ref{S:partition}, whilst the computation of fluxes is presented in Section \ref{S:fluxes}. Section \ref{S:LSM} presents the LSM, while the temporal advancement with the fractional-step method is described in Section \ref{S:time}.

%%%%%%%%%%%%%%%%%%%%%%%%%%%
\subsection{Domain partitioning and communication} \label{S:partition}

\begin{figure}[h!]
\centering
\includegraphics[width=0.75\linewidth]{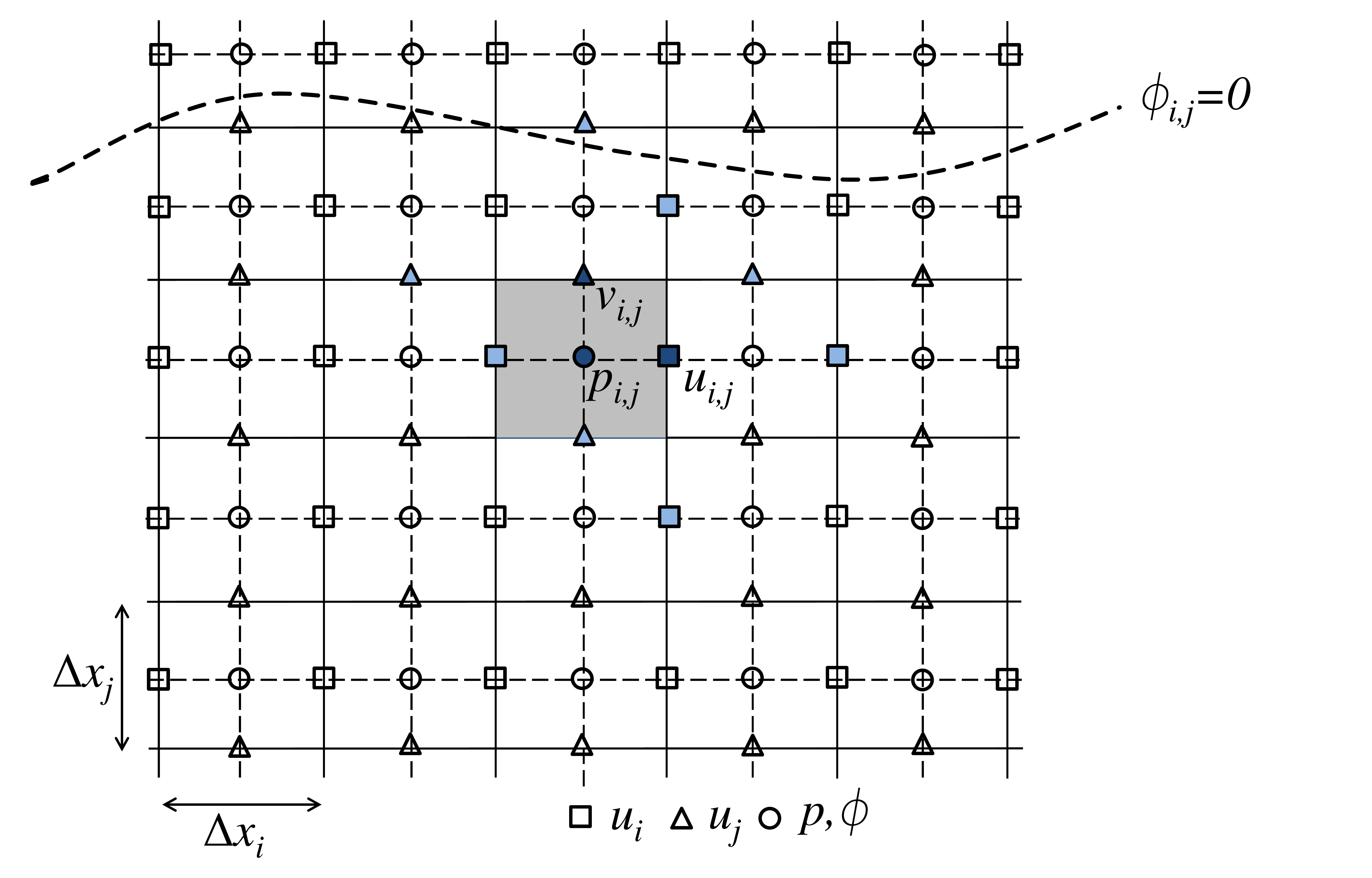}
\caption{Representation of the Cartesian grid and staggered velocity arrangement.} \label{fig:cartesian}
\end{figure}

$\mathtt{Hydro3D}$ adopts Cartesian meshes to discretise the computational domain. Figure \ref{fig:cartesian} presents the mesh distribution with uniform grid spacing in every spatial direction with pressure computed on the cell centres and velocities stored in a staggered fashion, i.e. on the cell faces. 
The grid spacing is $\Delta x_i = {x}_{i+1} - {x}_i$, and $\phi_{ij}$ represents the LSM function that is equal to zero at the two-phase fluid interface, as described later in Section \ref{S:LSM}.

Structured rectangular grids avoid the need for building and communicating connectivity matrices required to identify the neighbours of every cell in order to compute the fluxes.
%, thus enabling the use of highly-accurate numerical schemes with a reduced to compute fast the fluxes. 
For instance, velocity derivative at a cell $i$ using 2$^{nd}$- or 4$^{th}$-order centred finite differences use velocity values at $\{i-1,i+1\}$ or $\{i-2,i-1,i+1,i+2\}$, respectively. 
These schemes allow the adoption of a balanced domain decomposition of the numerical domain with an even number of grid cells per sub-domain, in which $n_i \times n_j \times n_k = n_t$ sub-domains are mapped onto $n_p$ processing units. %Usually one CPU has one sub-domain assigned so $n_t$ = $n_p$, unless stated otherwise. %UNAI: Esta ultima frase no me convence

The communication protocol between sub-domains is performed using MPI in which a number of halo or ghost cells, $n_g$ are used to exchange the required information, such as velocities or pressure values, between adjacent, overlapping sub-domains. 
The value of $n_g$ is set according to the stencil of the discretisation scheme to calculate the fluxes, being equal to the number of cells per direction plus one, i.e. $n_g$ = 2 and 3 for the 2$^{nd}$- or 4$^{th}$-order centred finite differences and 4 when using the 5$^{th}$-order Weighted Essentially Non-Oscillatory (WENO) scheme.

\begin{figure}[h!]
\centering
\includegraphics[width=0.8\linewidth]{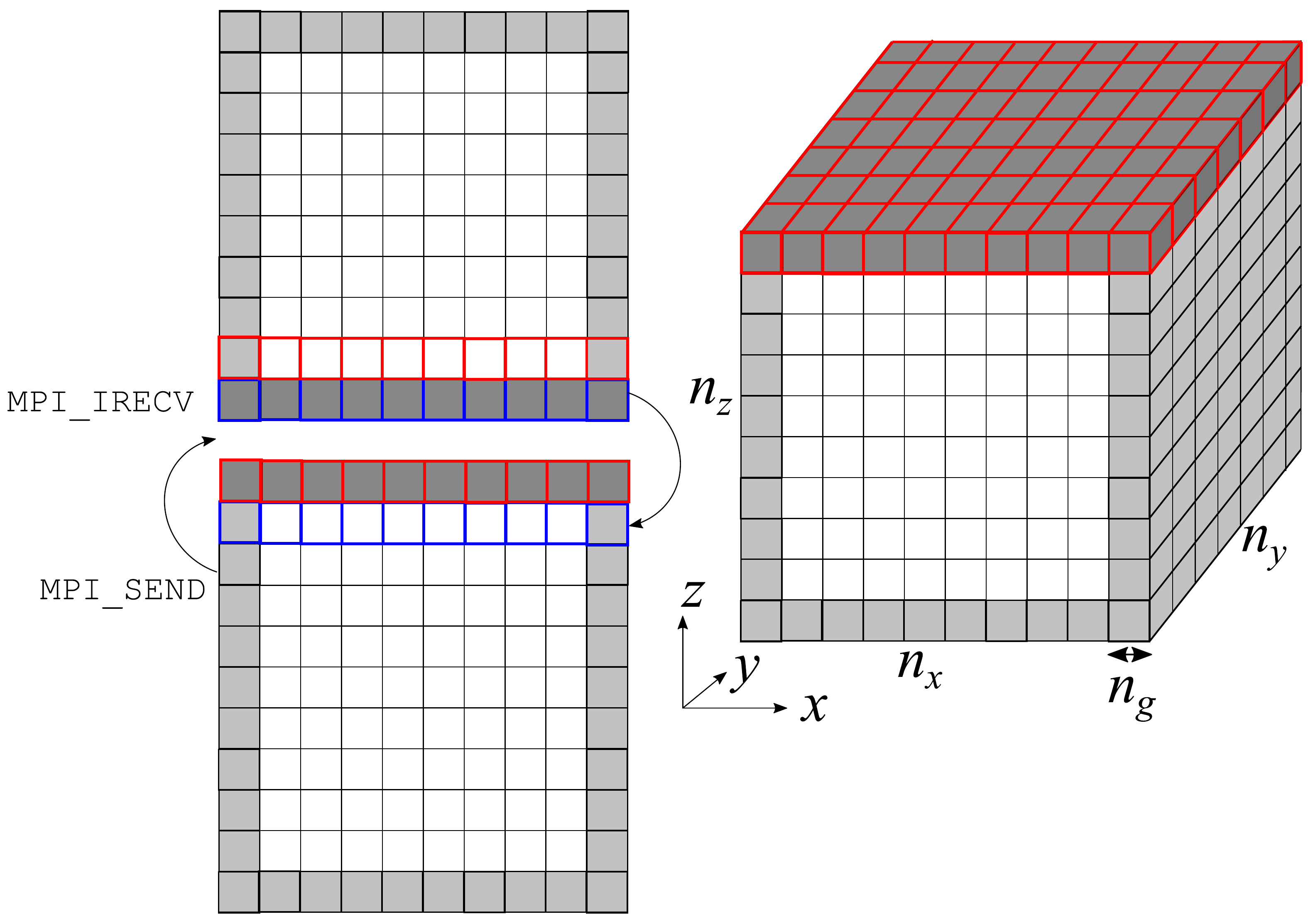}
\caption{Communication between sub-domains using MPI.} \label{fig:commMPI}
\end{figure}

The communication overhead when exchanging information by overlapping sub-domains is proportional to the number of ghost cells and sub-domain divisions over the considered \emph{ij} direction, i.e. $n_{GC}$ = $n_g \times n_{x_i} \times n_{x_j}$, as seen in Figure \ref{fig:commMPI} which depicts the process of data transfer between two adjacent sub-domains.
Hence, in cases with a reduced sub-domain division, there is a small number of large-size messages to be communicated. In contrast, when using a large number of sub-domains, communication is performed by a large number of small-sized messages.
%The latter can lead to a sub-optimal performance of the code.
In this version of $\mathtt{Hydro3D}$, point-to-point communications are achieved using a sequence of non-blocking \texttt{MPI\_IRECV} and blocking \texttt{MPI\_SEND} and \texttt{MPI\_WAIT} primitives.

%%%%%%%%%%%%%%%%%5
\subsection{Computation of fluxes} \label{S:fluxes}

The computation of the convective fluxes is the most important component in the discretisation of the numerical solution (\citet{Breuer1998}). 
The non-linear convective flux, $\mathcal{C} = u_j \partial_{x_j} u_i$, in the r.h.s of Eq. \ref{eq:ns2} can be approximated in $\mathtt{Hydro3D}$ using 4$^{th}$-order Central Differences (CD) or 5$^{th}$-order WENO schemes. 
Whilst CD schemes compute the fluxes from a single, centred stencil, 5$^{th}$-order WENO schemes perform a number of operations to find the optimum stencils which, in turn, increase the computational expense of the simulations. 
CD schemes are non-dissipative but can lead to instabilities in the simulations, as in the case of shock-waves. WENO schemes introduce numerical dissipation to provide a more stable solution that needs to be controlled to avoid excessive damping. 
Due to the inherent advantages of each scheme, hybrid CD-WENO schemes can be developed for simulations with shock-turbulence interaction \cite{Costa2007,Fernandez-Fidalgo2018}, but these are not in the scope of this paper.

In highly-accurate CD schemes a $p$-order in spatial accuracy is obtained by approximating velocity derivatives using kernels of, at least, $p+1$ cells in every spatial direction. 
Consequently, an increase in scheme accuracy leads to larger computing times, which can be cumbersome in DNS or LES, and thus a motivation to study the performance of $\mathtt{Hydro3D}$ when using each of these.

%% {Central differences schemes}
In a 2$^{nd}$-order CD scheme, the velocity gradient at a given cell results from the linear interpolation of velocities from the immediately adjacent cells, as

\begin{align}
    & f_i = \frac{\partial \textit{$u_i$}}{\partial x_i} =
    \frac{\textit{{u}}_{i+1} - \textit{{u}}_{i-1}}{2 \Delta x_i} \label{eq:fi}
\end{align}

The velocity derivative computed with 4$^{th}$-order CD fits at the $i$-cell a third-order polynomial through the adjacent four cells, which reads:

\begin{align}
    & f_i = \frac{-\textit{{u}}_{i+2} + 9\textit{{u}}_{i+1} - 9\textit{{u}}_{i-1} +\textit{{u}}_{i-2} }{16 \Delta x_i} \label{eq:4cds}
\end{align}

%% {WENO scheme}
WENO schemes are appropriate to deal with sharp gradients in the velocity fields by introducing artificial dissipation as an upwind scheme \cite{Shu1989}. 
This results in a more suitable approach for the computation of free-surface flows using LSM \cite{McSherry2018} as small discontinuities can appear at the interface. 
The basis of the 5$^{th}$-order WENO scheme is to build a high-order reconstruction of the derivative of the flux at a grid cell $x_i$, which requires a flux splitting to compute the positive and negative interface fluxes on the $x_i$-direction, $f_{i \pm 1/2}$, at the cell interfaces, i.e. $\Delta x_{i \pm 1/2}$, from three candidate stencils.
In the calculation of the velocity derivative $f_i$ with WENO, the optimal weights are: $c_1$ = 1/10, $c_2$ = 3/10 and $c_3$ = 6/10. The smoothness indicators, $\beta_k$, serve to scale the optimal weights and are calculated as:

\begin{align}
\beta_1 &= \frac{13}{12}(f_{i-2}-2f_{i-1}+f_i)^2 + \frac{1}{4}(f_{i-2}-4f_{i-1}+3f_i)^2  \label{eq:wenobeta1}  \\ 
\beta_2 &= \frac{13}{12}(f_{i-1}-2f_{i}+f_{i+1})^2 + \frac{1}{4}(f_{i-1}+f_{i+1})^2   \\ 
\beta_3 &= \frac{13}{12}(f_{i}-2f_{i+1}+f_{i+2})^2 + \frac{1}{4}(3f_{i}-4f_{i+1}+f_{i+2})^2  \\
\alpha_k &= \frac{c_k}{(\beta_k+\varepsilon_1)^m}    \hspace*{2cm}  k \text{ = 1, 2, 3}
\end{align}

Here $\varepsilon_1$ is added to avoid division by zero and set to a value equal to 10$^{-6}$, $m$ is set to 2, and weights $\alpha_k$ are normalised to guarantee convexity. Thereafter, the non-linear weights $\omega_k$ are obtained as,

\begin{align}
\omega_k = \frac{\alpha_k}{\sum_k \alpha_k}      \hspace*{2cm}  k \text{ = 1, 2, 3}
\end{align}

Finally, the velocity derivative at the cell $i$ is computed as:

\begin{align}
\begin{split}
f_{i} & =  \frac{1}{3}\omega_1f_{i-2} + \frac{1}{6}(7\omega_1+\omega_2)f_{i-1} + \frac{1}{6}(11\omega_1+5\omega_2+2\omega_3)f_{i}  \\ 
& +\frac{1}{6}(2\omega_2+5\omega_3)f_{i+1} - \frac{1}{6}\omega_3f_{i+2}  \label{eq:wenofi}
\end{split} 
\end{align}

Note that the latter flux reconstruction process is performed in the three spatial directions, i.e. $f_{{i\pm 1/2},j,k}$, $f_{{i},{j\pm 1/2},k}$ $f_{{i},j,{k\pm 1/2}}$, for each of the quantities considered, e.g. $u_i$.

In $\mathtt{Hydro3D}$, the linear diffusive or viscous, fluxes, $\mathcal{D} = \partial_j^2 u_i$, are computed with a 4$^{th}$-order CD scheme independently to the scheme used for the convective terms as these are of elliptic nature, which in its one-dimensional version is

%\begin{align}
%\mathcal{D} = & \nu \Delta t \left(
%\frac{\textit{{u}}_{i,j,k}}{\Delta x^2 + \Delta y^2 +\Delta z^2} - 
%-\frac{\textit{{u}}_{i+1,j,k} + \textit{{u}}_{i-1,j,k}}{\Delta x^2} \\ 
%& -\frac{\textit{{u}}_{i,j+1,k} + \textit{{u}}_{i,j-1,k}}{\Delta y^2} 
%-\frac{\textit{{u}}_{i,j,k+1} + \textit{{u}}_{i,j,k-1}}{\Delta z^2}      
%\right)
%\label{eq:diff2nd}
%\end{align}

\begin{align}
\mathcal{D}_i = & \nu \Delta t  \left( \frac{-\textit{{u}}_{i+2} + 16\textit{{u}}_{i+1} - 30\textit{{u}}_{i}  + 16\textit{{u}}_{i-1} -\textit{{u}}_{i-2}}{12 \Delta x_i^2}  \right)
\label{eq:diff4th}    
\end{align}

%%%%%%%%%%%%%
\subsection{Level-set method} \label{S:LSM}

$\mathtt{Hydro3D}$ adopts the Level-Set Method (LSM) to resolve Eulerian-Eulerian multi-phase flows defining a continuous level-set function, $\phi$, to determine the fluid density and viscosity fields across the computational domain, which was validated in \citet{Kara2015,Chua2019} and \citet{McSherry2018}.
The level-set function assigns positive or negative values depending on which fluid occupies every grid cell. 
In the present study, values of $\phi \geq 0$ denote water whilst $\phi \leq 0$ is air, with the air-water interface being defined by $\phi=0$ \cite{Kara2015} (see Figure \ref{fig:cartesian}), and a Heaviside function $\mathcal{H}(\phi)$ is used to smooth the transition of the density ($\rho(\phi)$) and dynamic viscosity ($\mu(\phi)$) fields across a layer of thickness $\varepsilon = 1.5 \Delta x_i$, as:

\begin{align}
& \rho (\phi) = \rho_a + (\rho_w - \rho_a ) \mathcal{H}(\phi) \label{eq:rho} \\
& \mu (\phi) = \mu_a + (\mu_w - \mu_a ) \mathcal{H}(\phi) \label{eq:mu} 
\end{align}

Here, sub-indices $a$ and $w$ denote values corresponding to the air and water, respectively. The Heaviside function $\mathcal{H}(\phi)$ defined as:

\[
\mathcal{H}(\phi) = 
\begin{cases}
0,                       & \text{\hspace{1cm} $\phi < -\varepsilon$} \\
\frac{1}{2} \left[1 + \frac{\phi}{\varepsilon} + \frac{1}{\pi} \text{sin} \left( \frac{\pi \phi}{\varepsilon} \right) \right], &     \text{\hspace{1cm} $\mathopen| \phi \mathclose| < \varepsilon$} \\
1,                      & \text{\hspace{1cm} $\phi > \varepsilon$} 
\end{cases}
\]

Free-surface flows resolved with LSM pose an additional computational overhead in comparison to the standard rigid-lid approach in which the air-water interface is fixed as the vertical upper boundary condition.
LSM requires resolving a non-linear hyperbolic advection equation in addition to the mass and momentum conservation equations, to account for the transport of $\phi$:

\begin{align}
 & \frac{\partial \phi}{\partial t} + \textit{u}_i \frac{\partial \phi}{\partial x_i} = 0 \label{eq:lsm}
\end{align}

This equation is resolved computing the derivatives with the 5$^{th}$-order WENO scheme and advanced in time using a third-order Total-Variation Diminishing Runge-Kutta (TVD-RK3) scheme, which is

\begin{align}
& \phi^k = \phi(x_i,t)  \\
& \phi^{k+1} = \phi^k - \Delta t \left( u_i \frac{\partial \phi^k}{\partial x_i}   \right) \label{eq:tv1} \\
& \phi^{k+2} = \frac{3}{4} \phi^k + \frac{1}{2} \phi^{k+1} - \frac{1}{4} \Delta t \left(u_i \frac{\partial \phi^{k+1}}{\partial x_i} \right) \label{eq:tv2} \\
& \phi^{k+3} = \frac{1}{3} \phi^k + \frac{2}{3} \phi^{k+1} - \frac{2}{3} \Delta t \left(u_i \frac{\partial \phi^{k+2}}{\partial x_i} \right) \label{eq:tv3}
\end{align}

Note that flux splitting is applied to compute each gradient of $\phi^{k}$, $\phi^{k+1}$ and $\phi^{k+2}$ at $\Delta x_{i\pm 1/2}$, which implies that for Eq. \ref{eq:tv1} six WENO fluxes for $\phi^{k}$ are calculated, i.e. $f_{{i\pm 1/2},j,k}$, $f_{{i},{j\pm 1/2},k}$ $f_{{i},j,{k\pm 1/2}}$, which is then repeated for $\phi^{k+1}$ and $\phi^{k+2}$. 

Due to the intrinsic nature of the $\phi$-advection equation, mass conservation is not guaranteed \cite{Sussman1994} as the required numerical stability requirement of $|\nabla \phi|=1$ is not automatically fulfilled.
This implies that a re-initialisation technique is computed after Eq. \ref{eq:lsm} \cite{Sussman1994}, which requires resolving another advection equation, Eq. \ref{eq:reini}, for the signed-distance function, $d_0$, solved with the TVD-RK3 scheme (Eq. \ref{eq:tv1} -- \ref{eq:tv3}) according to an "artificial time step" $\tau$ = CFL\textsubscript{LSM}$ \cdot \texttt{MAX}(\Delta x_i)$.
The smoothed signed function, $s(d_0)$, in Eq. \ref{eq:reini} is the calculated according to Eq. \ref{eq:reini2} adopting the initial condition $d_0(x_i,0)$ = $\phi(x_i,t)$. 

\begin{align}
& \frac{\partial d_0}{\partial \tau} = s(d_0) (1 - |\nabla d_0|) \label{eq:reini}    \\
& s(d_0) = \frac{d_0}{\left( d_0^2 + \left( |\nabla d_0| \varepsilon_r \right)^2  \right)^{0.5}}    \label{eq:reini2}
\end{align}

This re-initialisation process is iteratively solved until the condition of $|\nabla \phi|=1$ is partially satisfied up to a residual $\varepsilon_{LS}$ set to 5 $\cdot$ 10$^{-3}$. 
Additionally, a maximum of 15 iterations is set, since the free-surface layer in turbulent flows will experience small high-frequency oscillations due to the turbulent structures \cite{McSherry2018,Kang2012a}. These challenge fast numerical convergence. 
Identically to the $\phi$-equation resolution, 18 stencil reconstructions are required at every re-initialisation step in addition to the calculation of Eq. \ref{eq:reini}. 
This can notably increase the computational cost if the number of iterations required during the re-initialisation is large.

Alg. \ref{al:lsm} presents the procedure to resolve the LSM, with \texttt{T\textsubscript{LS}{}} and \texttt{T\textsubscript{LS\_$\omega_k$}{}} labelling the time spent on solving the whole LSM and that only for the WENO weights in LSM, respectively.

\begin{algorithm}[ht!]
  \caption{Pseudocode for advection of $\phi$ function in LSM}\label{al:lsm}
  \begin{algorithmic}[1]
\State \texttt{$\phi^t = \phi(x_i,t)$} \Comment{\texttt{T\textsubscript{LS}{}=MPI\_WTIME}}
    \For{k = 1, 3}            
       \State \texttt{WENO stencil for $\phi^t$}  \Comment{ \texttt{T\textsubscript{LS}{}} \texttt{\textsubscript{$\omega_k$}{}}}
       \State \texttt{Compute $\phi^{t+k}$ with Eqs. \ref{eq:tv1} - \ref{eq:tv3}}
    \EndFor
\State{\texttt{Re-initialisation process}}
\For{m = 1, 15} 
    \State \texttt{$d_0 = \phi^{t+1}$}
    \While{$|\nabla d_0| - 1 { > } \hspace*{1pt} \varepsilon_{LS}$ and m $\geq$ 2}
    \For{k = 1, 3}  
       \State \texttt{WENO stencil for $d_0^{k}$}   \Comment{ \texttt{T\textsubscript{LS}{}} \texttt{\textsubscript{$\omega_k$}{}}}
       \State \texttt{Compute $|\nabla d_0^{k}|$}
       \State \texttt{Compute $s(d_0^{k})$}
       \State \texttt{Compute $d^{k}$ with Eqs. \ref{eq:tv1} - \ref{eq:tv2}}
    \EndFor
      \State {call \texttt{MPI\_ALLREDUCE($|\nabla d_0|$)}}
       \EndWhile  
    \EndFor 
\State \texttt{$\phi^{t+1} = d_0^{k}$}
\State \texttt{Heaviside function: $\mathcal{H}(\phi^{t+1})$}
\State \texttt{Update $\rho(\phi^{t+1})$ and $\mu(\phi^{t+1})$ fields} \Comment{\texttt{T\textsubscript{LS}{}=T\textsubscript{LS}{}-MPI\_WTIME}}   
\end{algorithmic}
\end{algorithm} 

Note that the computational expense of using LSM in turbulent flows is intrinsically high due to the fine grid resolution needed to capture the instabilities at the air-water interface. Too coarse a grid resolution will challenge numerical stability. 
For this reason, WENO is used to compute both the LSM advection equations and convective fluxes in the momentum equation. 
%This scheme is restricted to the code's fixed rectangular Cartesian grid nature.
New insights into the computational cost of using LSM for free-surface flows is of special relevance for the engineering research community, defining how different HPC chip architectures impact on the performance of computing LSM.

%%%%%%%%%%%%%%%%%%%%%5
\subsection{Temporal advancement} \label{S:time}

The standard fractional-step method \cite{Chorin1968} is used to advance the equations in time, as its step-wise nature allows the development of highly-scalable parallelised codes. 
The adopted fractional-step method is fully explicit and divides the computing sequence at each time step into three main stages as depicted from Alg. \ref{al:fs}.
First, the predicted non-divergence-free velocity field, $u_i^*$, is calculated from the pressure and velocities obtained at the previous time step, $p^t$ and $u_i^t$, accomplished using an explicit time advancement with a three-stage Runge-Kutta method. The second step comprises the resolution of the Poisson pressure equation. This is iteratively solved using a multi-grid method in order to correct the predicted velocity field until the divergence-free condition, i.e. $\nabla \cdot u^*_i$, is below a given residual $\varepsilon$ equal to $10^{-6}$. 
In the final step the velocity and pressure are updated.
If the LSM method is used, this is computed before the velocity fluxes.

$\mathtt{Hydro3D}$ has an internal profiler to track the time spent in the main subroutines at every time step, obtained with the master processor and using the \texttt{MPI\_WTIME} command.
These are highlighted in Alg. \ref{al:fs} and account for:

\begin{itemize}
    \item \texttt{T\textsubscript{TT}{}}: Total time per time step or mean runtime.
    \item \texttt{T\textsubscript{LS}{}}: Level-Set Method.
    \item \texttt{T\textsubscript{CD}{}}: Convective and viscous fluxes computed over the three-step Runge-Kutta (RK) to obtain the non-divergence free velocity.
    \item \texttt{T\textsubscript{P}{}}: Resolution of the Poisson pressure equation.
    \item \texttt{T\textsubscript{up}{}}: Update of the velocity and pressure fields. 
\end{itemize}

Note that subroutines consuming a reduced part of the overall computing time per time step, e.g. less than 1.0\%, are normally excluded from the analysis, e.g. sub-grid scale model or time step calculation based on the CFL condition, and thus not presented in the results.

In terms of collective communications, two main \texttt{MPI\_ALLREDUCE} calls are performed to determine velocity maxima that condition the time advancement of the simulation. 
First, the maximum velocity from the previous time step, $u_i^t$, is fetched to determine the time step $\Delta t$ according to the adopted CFL value. 
Then the divergence free condition in the Poisson pressure solver is checked with the non-solenoidal velocity $u_i^*$ after every iteration of the solver (lines 17--19 in Alg. \ref{al:fs}), with the code needing to retrieve \texttt{MAX}$(\nabla \cdot u^*_i)$ from every sub-domain, at every pressure solver iteration.

\begin{algorithm}[ht!]
  \caption{Fractional-step method adopted in $\mathtt{Hydro3D}$}\label{al:fs}
  \begin{algorithmic}[1]
    \State \texttt{Variable allocation and initialisation}
\nonumber{\texttt{  }}
    \For{t = $t_0, t_{end}$}            \Comment{\texttt{T\textsubscript{TT}{}=MPI\_WTIME}}
           \State \texttt{Store $u^{t}_i = u_i^{t-1}$}
\nonumber{\texttt{  }}
        \If{\texttt{Level-Set Method}}  \Comment{\texttt{T\textsubscript{LS}{}=MPI\_WTIME}}
           \State \texttt{Alg. \ref{al:lsm}}
        \EndIf  \Comment{\texttt{T\textsubscript{LS}{}=T\textsubscript{LS}{}-MPI\_WTIME}}       
\nonumber{\texttt{  }}
    \For{$t_{RK}$ = 1, 3} \Comment{\texttt{T\textsubscript{CD}{}=MPI\_WTIME}}
    
\nonumber{\hspace{-1.3cm} \texttt{Convective fluxes,} $\mathcal{C}$ :}
        \If{\texttt{ $2^{nd}$-order CD}}
           \State \texttt{Eq. \ref{eq:fi}}
        \ElsIf{\texttt{ $4^{th}$-order CD}}   
           \State \texttt{Eq. \ref{eq:4cds}}
        \Else{\texttt{ WENO}}   
           \State \texttt{Eqs. \ref{eq:wenobeta1} - \ref{eq:wenofi} }
        \EndIf
        
        %\Comment{T\textsubscript{C}{}=T\textsubscript{C}{}-\texttt{MPI\_WTIME}}
        %\Comment{T\textsubscript{D}{}=\texttt{MPI\_WTIME}}
        
\nonumber{\hspace{-1.3cm} \texttt{Viscous fluxes,} $\mathcal{D}$:\hspace{.1cm} \texttt{Eq. \ref{eq:diff4th}}}
    \EndFor  \Comment{\texttt{T\textsubscript{CD}{}=T\textsubscript{D}{}-MPI\_WTIME}}
\nonumber{\texttt{  }}

\nonumber{\hspace{-1.3cm} \texttt{Non-divergence free velocity: }}

%\State \texttt{Determine source term $f_i$}
\State ${u_i}^*$ = ${u_i}^{t}$ + $\Delta t$ { }($\mathcal{D}$ $({u_i}^{t}$) + $\mathcal{C}$ (${u_i}^{t}$) + $\nabla p^t$ + $S_i$) %\Comment{T\textsubscript{CD}{}=T\textsubscript{CD}{}-\texttt{MPI\_WTIME}}
% \nonumber{\texttt{  }}
% \State \texttt{Computation of the n}

\nonumber{\hspace{-1.3cm} \texttt{Solve Poisson pressure equation: }\Comment{\texttt{T\textsubscript{P}{}=MPI\_WTIME}}}
       \While{  $\nabla \cdot u_i^* {  >  } \hspace*{1pt} \varepsilon$   }
       \State \texttt{$\nabla^2 \hat{p} = ({\nabla \cdot u_i^* })/{\Delta t}$}
       \EndWhile \Comment{\texttt{T\textsubscript{P}{}=T\textsubscript{P}{}-MPI\_WTIME}}

\nonumber{\hspace{-1.3cm} \texttt{Update velocity and pressure fields: } \Comment{\texttt{T\textsubscript{up}{}=MPI\_WTIME}}}
    \State \texttt{    $u^{t+1} = u_i^* + \Delta t \nabla \hat{p}$ }
    \State \texttt{    $p^{t+1} = p^t + \hat{p} - {\nu \Delta t \Delta \hat{p}}/{2}$ } 
    \Comment{\texttt{T\textsubscript{up}{}=T\textsubscript{up}{}-MPI\_WTIME}}    

\nonumber{\texttt{  }} \Comment{\texttt{T\textsubscript{TT}{}=T\textsubscript{TT}{}-MPI\_WTIME}}  

\nonumber{\hspace{-1.3cm} \texttt{Time averaging and write output files}}
    \EndFor    
  \end{algorithmic}
\end{algorithm}

%\begin{align}
%&\frac{\tilde{\textbf{u}}-\textbf{u}^{l-1} }{{\Delta t} }=
%  \alpha_l (\nu+\nu_t) \nabla^2 \textbf{u}^{l-1} - \alpha_l \nabla p^{l-1} 
%- \alpha_l \mathtt{C}^{l-1}   %- \alpha_l [\textbf{u}(\nabla \cdot \textbf{u})]^{l-1}   
%- \beta_l \mathtt{C}^{l-2}    % [\textbf{u}(\nabla \cdot  \textbf{u})]^{l-2}   		  
%\label{eq:frac1}\\
%& \tilde{\textbf{u}}^{*}=\tilde{\textbf{u}} + \textit{\textbf{f}} \Delta t \label{eq:step4}\\
%& \nabla^2 \tilde{p} = \frac{\nabla \cdot \tilde{\textbf{u}}^{*}}{\alpha_l \Delta t} \label{eq:frac2}\\
%& \textbf{u}^{t} = \tilde{\textbf{u}}^{*} - \alpha_l \Delta t \nabla\tilde{p} \label{eq:frac3}\\
% & p^{t} = p^{t-1} + \tilde{p} - \frac{\alpha_l \Delta t}{2 Re} \nabla^2 \tilde{p} \label{eq:frac4} 
%\phantom{\hspace{0cm}} %%<---adjust the value as you want
%\end{align}
%\begin{align}
%    & \nabla^2 \hat{p} = \frac{\nabla \cdot u_i }{\Delta t} \label{eq:poisson}
%\end{align}
%\begin{align}
%    & u_{t+1} = u_t + \Delta t \cdot \nabla \hat{p} \label{eq:upvel} \\
%    & p_{t+1} = p_t + \hat{p} - \frac{\nu \Delta t \Delta \hat{p}}{2} \label{eq:uppre}
%\end{align}
%%%%%%%%%%%%%%%%%%%%%%%%%%%%%%%5
%%%%%%%%%%%%%%%%%%%%

%\newpage
%\section{Experimental setup} 
\section{Cluster architectures} \label{S:3}

In this work, three HPC systems are used, each equipped with distinctive dual-processor computing nodes. Again, for convenience, these systems are codenamed as {EPYC}, {SKL} and {TX2}, which use
AMD EPYC Rome, Intel Skylake and ARM Marvell ThunderX2 processors respectively. Full system descriptions are provided in Table \ref{tab:systems}.

\begin{table}[h]
\caption{Description of the three HPC systems.}  \label{tab:systems}
\begin{footnotesize}
\begin{tabular}{|l|l|l|l|}
\hline
\textbf{System name}        & {ThunderX2}    &  {Skylake}     &  {EPYC-Rome}  \\ \hline
\textbf{Computing Service}     & GW4 Isambard          & Supercomputing        & Supercomputing  \\ 
                            &                       & Wales (Hawk)          & Wales (Hawk) \\ \hline
\textbf{Number of nodes}    & 128                   & 201                   & 64    \\ \hline
\textbf{Core count/node}  & 2 CPUs$\times$32 = 64 & 2 CPUs$\times$20 = 40 & 2 CPUs$\times$32 = 64 \\ \hline
\textbf{Memory}             & 256 GB DDR4           & 192 GB DDR4           & 256 GB DDR4\\
\textbf{per node}           & @2400MHz              & @2666MHz              & @2500MHz\\ \hline
\textbf{Interconnect}       & Cray Aries            & Infiniband EDR        & Infiniband EDR    \\ \hline
\textbf{\it{CPU spec:}}     &                       &                       &  \\ \hline
\textbf{Architecture}   & ARM v8.1              & Intel Skylake         & Rome (Zen 2)   \\ \hline
\textbf{Model}          & Marvell ThunderX2     & Intel Xeon 6148 Gold  & EPYC-Rome 7502 \\ \hline

\textbf{Clock}          & 2.10 GHz              & 2.40 GHz              & 2.50 GHz     \\ \hline
\textbf{Memory channels}    & 8                     & 6                     & 8    \\ \hline
%\textbf{Bandwidth}          & 320 GB/s              & 256 GB/s              & 400 GB/s     \\ \hline
\textbf{TDP}                & 175 W                 & 150 W                 & 180 W    \\ \hline

\end{tabular} \end{footnotesize}
\end{table}

The TX2 system is the full ARM partition of Isambard, the first large-scale ARM-based production level HPC system in the UK managed by the GW4 Alliance, comprising the Universities of Bristol, Bath, Cardiff and Exeter, together with Cray and hosted at the UK Met Office. 
Isambard is a Cray XC50 system containing over 10,000 ARM cores and a full Cray software stack. More information about the system, including the full specification, is available at \footnote{GW4 Isambard - https://gw4.ac.uk/isambard}.

The SKL and EPYC systems are partitions of Hawk, an HPC system hosted by Supercomputing Wales (SCW) at Cardiff University. SCW is a pan-Wales UK project featuring the Universities of Cardiff, Swansea, Bangor and Aberystwyth. 
At the time of writing, the SKL partition comprises 201 dual processor nodes, totalling 8,040 cores and 46 TByte of memory, whilst the 64 node EPYC partition comprises 4096 EPYC-Rome 7502 cores with 50 TByte of memory. 
More information about the system and the project is available at \footnote{Supercomputing Wales - https://www.supercomputing.wales}.

%%%%%%%%%%%%%%%%%%%%%%%%%%%%%%%%%%%%%%%%%%%
\section{Benchmarks description} \label{S:4}

This section describes the three benchmarks selected to analyse the performance of $\mathtt{Hydro3D}$, designed to deliver a range of configurations that provide a varying balance of compute- and memory-bound simulations.

%Note that in the lid-driven cavity flow three ghost cells are used while in the other benchmarks the number of ghost layers increases to four as they adopt a discretisation scheme of one order of accuracy higher.

\subsection{3D lid-driven cavity flow}

The first benchmark comprises a three-dimensional (3D) lid-driven cavity flow simulated using 4$^{th}$-order CD to compute the convective fluxes, which is the fastest compute discretisation scheme using Cartesian grids. 
This minimises the overhead in computing convective and diffusive fluxes. 
In order to analyse the code performance, six problem sizes with increasing number of grid cells are adopted. 

The computational domain is a cube with dimensions [0,1]$\times$[0,1]$\times$[0,1] and an inlet velocity of $U_0(x_i)$ = (0.0,0.0,1.0) set at the top boundary, i.e. at $z=1$, as a slip condition. No-slip conditions are used at $x = 0$ and 1 and at the cube bottom, $z=0$, whilst periodic conditions are set at the transverse faces at $y=0$ and 1. The mesh resolution is uniform in the three spatial directions and variable time step is set with a CFL condition equal to 0.8. The Reynolds number is set to 400 as adopted in previous scalability studies performed by Ouro et al. \cite{Ouro2019CAF}, who previously validated the velocity field with $\mathtt{Hydro3D}$.

\subsection{Taylor-Green vortex}

The second benchmark comprises the simulation of the incompressible Taylor-Green Vortex (TGV) adopting the 5$^{th}$-order WENO scheme for the calculation of the convective terms. 
This case is often selected to evaluate the numerical properties of high-order discretisation schemes, e.g. scheme dissipation rate \cite{Fehn2018,Fernandez-Fidalgo2020}, such as the 5$^{th}$-order WENO. 
The case simulated has a $Re$ = $L U_0/ \nu$ = 1,600 with $U_0$ equal to 1.0 and characteristic spatial scale $L$ = 1 as the cube's edge length. The spatial domain extends $0 < x,y,z < L$ with periodic boundary conditions on the three spatial directions. 
A variable time step was set with a CFL condition of 0.3. 
The initial flow field is defined according to the following velocity distribution:

\begin{align}
    & u(x,y,z) = U_0 \text{ sin } \left(\frac{x}{L}\right) \text{ cos } \left(\frac{y}{L}\right) \text{ cos } \left(\frac{z}{L}\right) \\
    & v(x,y,z) = - U_0 \text{ cos } \left(\frac{x}{L}\right) \text{ sin } \left(\frac{y}{L}\right) \text{ cos } \left(\frac{z}{L}\right) \\
    & w(x,y,z) = 0 
\end{align}

\begin{figure}
\centering
\includegraphics[width=0.9\linewidth]{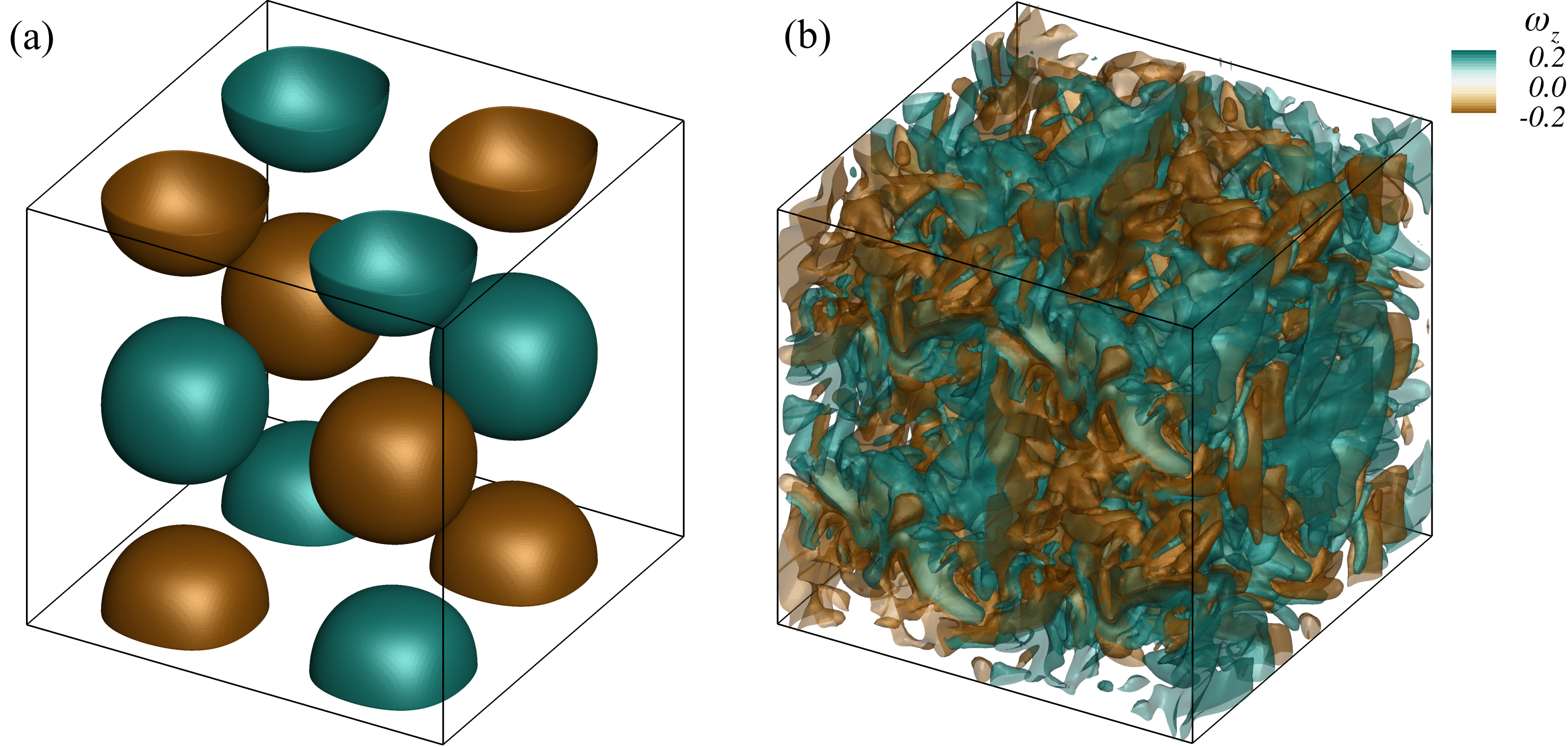}
\caption{Iso-surfaces of $z$-vorticity at non-dimensional times $t$ = 0 and 12.} \label{fig:tgv_isos}
\end{figure}

Figure \ref{fig:tgv_isos} presents iso-surfaces of $z$-vorticity ($\omega_z = \partial_x v - \partial_y u$) obtained at non-dimensional times $t$ = 0 and 12 using a computational mesh with 32.8 million elements. These depict how the initial large-scale vortices transition to smaller scale structures as the simulation advances in time.

\subsection{Solitary wave}

\begin{figure}[h!]
\centering
\includegraphics[width=0.99\linewidth]{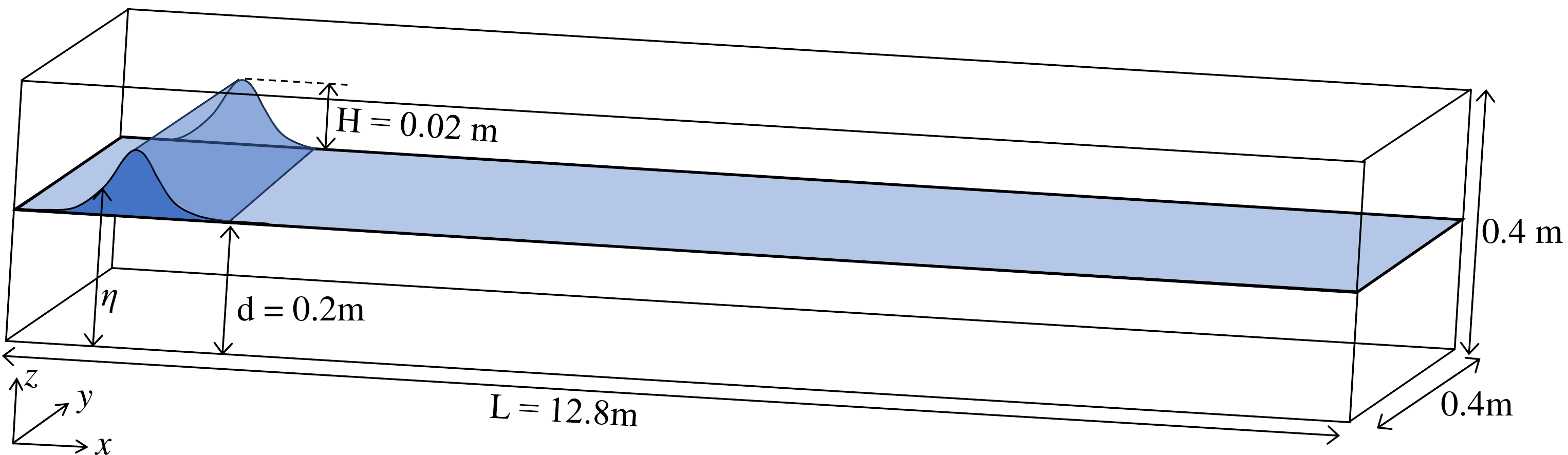}
\caption{Schematic of the solitary wave benchmark case.} \label{fig:wave_sketch}
\end{figure}

In the third benchmark, a solitary wave propagating across a numerical wave tank is simulated, which represents one of the current modelling challenges in computational hydraulics \cite{Rodi2017}. 
The air-water interface is resolved using the LSM which implies a higher computational load than the previous benchmarks, as indicated in Section \ref{S:LSM}. 
%Validation of a similar solitary wave case and other wave-structure interaction problems was presented by \citet{Christou2020}.
The numerical wave tank measures 12.8 m $\times$ 0.4m $\times$ 0.4m and is presented in Figure \ref{fig:wave_sketch}, with a water depth $d$ set to 0.2 m and solitary wave amplitude $H$ of 0.02 m. 
The density of the water is 1,000 kg m$^{-3}$ and that of air 1.25 kg m$^{-3}$, with viscosity values of 1$\cdot$10$^{-6}$ m$^2$ s$^{-1}$ and 1.8$\cdot$10$^{-5}$ m$^2$ s$^{-1}$, respectively. 
A constant time step is set to $\Delta {t}$ = 0.001 s, whilst for the advection equation the pseudo-time step ($\Delta \tau$) is variable with a CFL = 0.10 to guarantee numerical stability \cite{Kang2012a,Calderer2014}.

The solitary wave is generated using Boussinesq theory \cite{Lee1982} with Eq. \ref{eq:wave_eta} defining the wave elevation, $\eta$, and Eqs. \ref{eq:wave_kw} and \ref{eq:wave_c} denoting wavenumber, $k_w$, and wave celerity, $c$, respectively.

\begin{align}
& \eta (t) = H \text{cosh}^{-2} (k_w ( x - c t) ) \label{eq:wave_eta} \\
& k_w = \sqrt{{3 h}/{4 d^3}} \label{eq:wave_kw} \\
& c = \sqrt{g (H + d)} \label{eq:wave_c}
\end{align}

Here $g$ represents gravity acceleration equal to 9.81 m s$^{-2}$ and $t$ is time. At the start of the simulations, velocities are set to zero across the computational domain. Thereafter, considering the wave elevation is normalised as $\eta_h = \eta / H$ and $\varepsilon = H / d$, the velocities at the inflow condition ($x_{in} = (0,y,z)$) are prescribed according to:

\begin{align}
\begin{split}
& u(x_{in},t) = \varepsilon \sqrt{g d}  \left[ \eta_h - \frac{\varepsilon \eta_h^2}{4}  
 + \frac{d^2}{3 c^2} \left( 1 - \frac{3 z^2}{2 d^2} \right)\frac{\partial^2 \eta_h}{\partial t^2}\bigg\rvert_{t=0 } \right]
\end{split} \\
& v(x_{in},t) = 0 \\
\begin{split}
& w(x_{in},t) = z \frac{\varepsilon}{c} \sqrt{g d}  \left[ 
\left(1- \frac{\varepsilon \eta_h }{2} \right) \frac{\partial \eta_h}{\partial t} + \frac{d^2}{3c^2}
\left(1 - \frac{z^2}{2 d^2} \right) 
\frac{\partial^3 \eta_h}{\partial t^3}\bigg\rvert_{t=0 } \right]
\end{split}
\end{align}

%%%%%%%%%%%%%%%%%%
%\newpage
%\section{Single node performance} \label{S:5}

%%%%%%%%%%%%%%%%%%%%%%%%%%%%%%%%%%%%%%%%%%%%%%%%%%%%
%\pagebreak \newpage

\section{Results} \label{S:results}

This section presents the performance results of $\mathtt{Hydro3D}$ based on an analysis of strong scaling behaviour with increasing problem size.
The comparison between the three HPC systems is reported on a node-to-node basis, with each configuration designed to populate each node with a single MPI process per core. 
Table \ref{t:cases} lists the configurations adopted in terms of node count and sub-domain divisions which are equal to the number of cores used.
Note that the TX2 and EPYC nodes have 64 cores per node, whilst the SKL nodes have 40. 
Given resource availability, the simulations with SKL and EPYC nodes on the Hawk cluster are run on a maximum of 64 nodes, i.e. 2,560 and 4,096 cores respectively, whilst those with TX2 nodes on the Isambard cluster used up to 125 nodes, i.e., 8,000 cores. 

\begin{table}[ht!]
\caption{Configurations of the simulations carried out on the ThunderX2 (TX2), Skylake (SKL) and EPYC-Rome (EPYC) systems for the three benchmark cases, including the number of nodes used (N) and sub-domains ($n_t = n_x \times n_y \times  n_z$) that is equal to the number of cores.} 
\begin{footnotesize} 
\begin{tabular}{p{4pt}p{16pt}|p{3pt}p{16pt}|p{3pt}p{16pt}|p{3pt}p{16pt}|p{3pt}p{16pt}|p{3pt}p{16pt}|p{3pt}p{16pt}|p{3pt}p{16pt}|p{3pt}p{16pt}}
\multicolumn{6}{c}{Benchmark 1: Cavity flow}&  \multicolumn{6}{c}{Benchmark 2: TGV}&  \multicolumn{6}{c}{Benchmark 3: solitary wave}\\
\multicolumn{2}{c}{TX2}&\multicolumn{2}{c}{SKL}&\multicolumn{2}{c}{EPYC}&
\multicolumn{2}{c}{TX2}&\multicolumn{2}{c}{SKL}&\multicolumn{2}{c}{EPYC}&
\multicolumn{2}{c}{TX2}&\multicolumn{2}{c}{SKL}&\multicolumn{2}{c}{EPYC}\\
\hline
%Nodes & Cores & Nodes & Cores & Nodes & Cores & Nodes & Cores & Nodes & Cores & Nodes & Cores & Nodes & Cores & Nodes & Cores & Nodes & Cores \\
N & $n_t$ &N & $n_t$ &N & $n_t$ &N & $n_t$ &N & $n_t$ &N & $n_t$ &N & $n_t$ &N & $n_t$ &N & $n_t$\\
1   & 64   & 1  & 40   & 1  & 64   & 1  & 64   & 1  & 40   & 1  & 64   & 1  & 64   & 1  & 40   & 1  & 64   \\
2   & 128  & 2  & 80   & 2  & 128  & 2  & 128  & 2  & 80   & 2  & 128  & 2  & 128  & 2  & 80   & 2  & 128  \\
4   & 256  & 4  & 160  & 4  & 256  & 4  & 256  & 4  & 160  & 4  & 256  & 4  & 256  & 4  & 160  & 4  & 256  \\
8   & 512  & 8  & 320  & 8  & 512  & 8  & 512  & 8  & 320  & 8  & 512  & 8  & 512  & 8  & 320  & 8  & 512  \\
10  & 640  & 10 & 400  & 10 & 640  & 16 & 1000 & 16 & 640  & 16 & 1000 & 16 & 1024 & 16 & 640  & 16 & 1024 \\
20  & 1280 & 16 & 640  & 16 & 1024 & 25 & 1600 & 25 & 1000 & 25 & 1600 & 32 & 2048 & 32 & 1280 & 32 & 2048 \\
25  & 1600 & 25 & 1000 & 20 & 1280 & 32 & 2000 & 32 & 1280 & 32 & 2000 & 40 & 2560 & 64 & 2560 & 40 & 2560 \\
32  & 2000 & 32 & 1280 & 25 & 1600 & 63 & 4000 & 63 & 2500 & 63 & 4000 & 64 & 4096 &    &      & 64 & 4096 \\
63  & 4000 & 52 & 2048 & 32 & 2000 &    &      &    &      &    &      &    &      &    &      &    &      \\
125 & 8000 & 64 & 2560 & 63 & 4000 &    &      &    &      &    &      &    &      &    &      &    &      \\
\end{tabular}
\end{footnotesize}
\label{t:cases}
\end{table}

The code was compiled with Intel Compiler v2017.4 using the \texttt{-O2} flag for the EPYC and SKL platforms, and executed using the Intel MPI Library v2017.4. On the Isambard platform, the code was compiled using the Cray Compiling Environment (CCE) v9.1.3 with the \texttt{-O2} flag and executed with the Cray MPI library using the ALPS system v6.6.

All benchmark simulations are run for 50 time steps, with mean computing times calculated based on the \texttt{MPI\_WTIME} command averaged over the last 40 time steps,
%, over which the number of iterations in the multi-grid pressure solver corresponded to the minimum set, i.e. two iterations, and
over which the standard deviation of the computing times is lower than $0.1\%$.
Note that during the domain partitioning, the sub-domains size is kept uniform for all cases so as to avoid load imbalance between parallel processes.

The runtime values presented from the internal code profiler includes both computing and MPI operations. The breakdown of the main computing subroutines are calculated using the profilers CrayPat on Isambard (TX2) and TAU v2.27 \cite{TAUprofiler2006} on Supercomputing Wales (SKL and EPYC), of value in characterising how systems distribute the computing workload for the same setup. The breakdown of MPI directives is obtained on Hawk using the IPM profiler. %only in Hawk 

\subsection{Node characterisation}\label{S:6.0}

In order to gain a sense of the raw computing power of the HPC clusters, a set of stress benchmarks on individual nodes of each system were undertaken, including Firestarter v1.7.4 to stress CPU cores and gather their peak performance, and STREAM to measure memory bandwidth. 
These two benchmarks are described in \cite{6604507} and \cite{mccalpin1995memory}, respectively.

The Firestarter binary available in its official repository was used, while STREAM was compiled with the Intel Compiler v2018.4 for the EPYC and SKL systems. 
The STREAM array size was $2^{25}$, with the reported values corresponding to the TRIAD test. The performance figures for TX2 are those reported in \citet{8945642}, as Firestarter did not run on the system due to an incompatibility with ARMv8.

\begin{table}[h]
\centering
\caption{Measured peak performance and memory bandwidth for individual nodes. Results from Firestarter and STREAM benchmarks.}
\label{tab:single_n_benchmarks}
\begin{tabular}{l|r|r|r|}
\cline{2-4}
      & \text{ThunderX2}  & \text{Skylake} & \text{EPYC-Rome} \\ \hline
\multicolumn{1}{|l|}{TFLOP/s}   & 1.28     & 2.06   &  1.99     \\ \hline
\multicolumn{1}{|l|}{GB/s}      &  288.00  & 180.36 &  264.90               \\ \hline
\end{tabular}
\end{table}

Results are shown in Table \ref{tab:single_n_benchmarks} for the three systems. SKL and EPYC clusters deliver a similar level of TFLOP/s performance($\sim$3\% difference between them) whilst the EPYC system achieves 46.8\% greater memory bandwidth. 
TX2 shows the lowest TFLOP/s performance figure but delivers the greatest memory bandwidth amongst the three chips, i.e. 8.7\% and 59.7\% more than EPYC and SKL processors, respectively.

\subsection{Lid-driven cavity flow} \label{S:6.1}

Six grid resolutions are adopted for the first benchmark in order to perform a set of tests that gradually evolve from a computing-bound scenario at low node count to a communication-bound scenario at higher node count. Details of the computational grids are provided in Table \ref{t:cav_res} with spatial resolution ($\Delta {x_i}$), number of elements per spatial direction ($n_{x_i}$) and total number of elements comprising the entire problem ($N_e = n_{x_i}^3$) in millions. 

\begin{table}[h]
\centering
\caption{Numerical details of the lid-driven cavity flow cases simulated.}
\begin{tabular}{lcccc}
\hline
 $\Delta {x_i}$ & $n_{x_i}$ & N\textsubscript{e}{} ($\cdot 10^6$) \\
\hline
0.00625   & 160   & 4.1 \\
0.00500   & 200   & 8.0\\
0.003125  & 320   & 32.8\\
0.00250   & 400   & 64.0\\
0.00125   & 800   & 512.0\\
0.00100   & 1,000 & 1,000.0\\
\hline
\end{tabular}  \label{t:cav_res}
\end{table}

\begin{figure}[h!]
\centering
\includegraphics[width=0.85\linewidth]{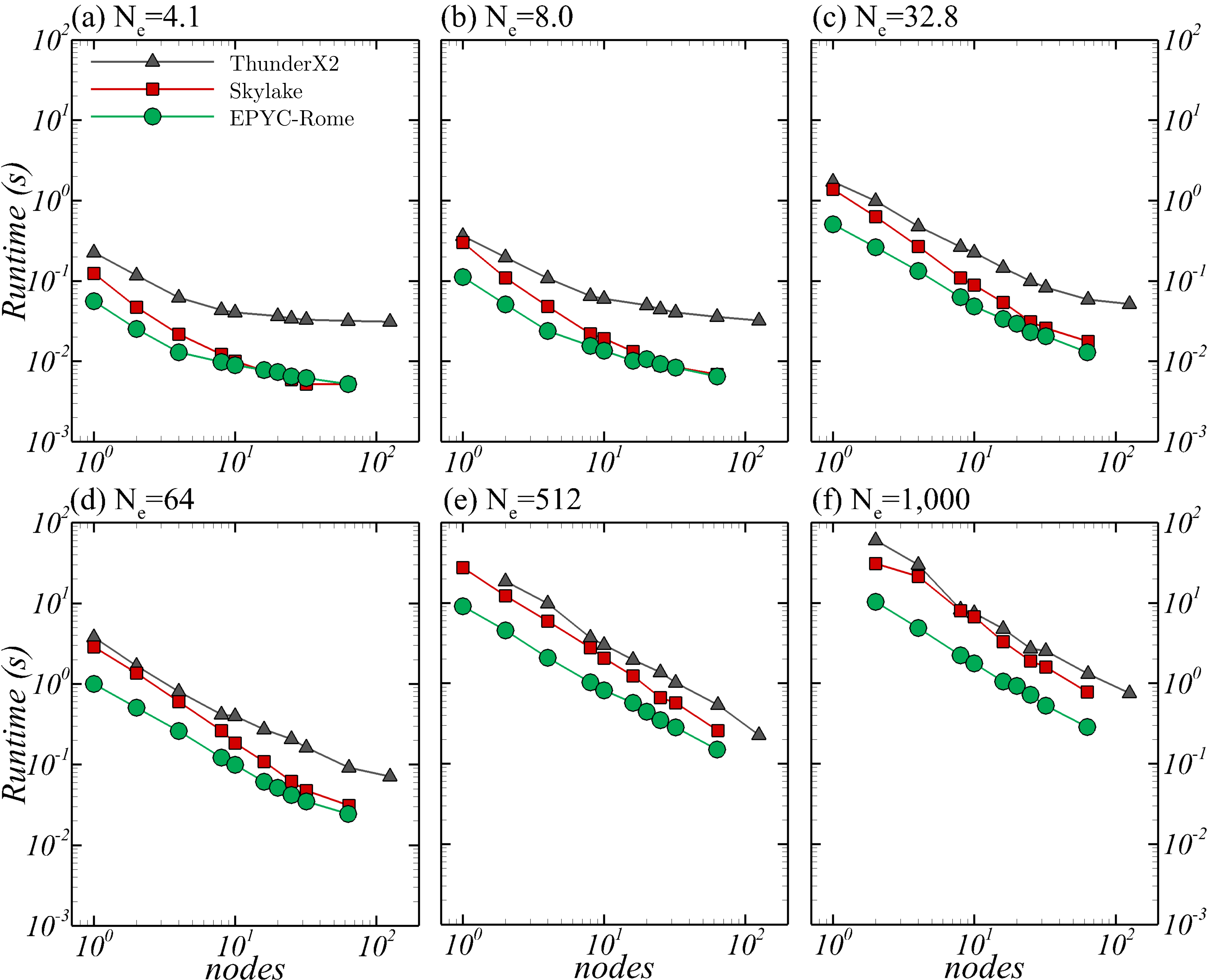}
\caption{Lid-driven Cavity flow: Results of averaged time per time step values in seconds for the three HPC architectures and six grid sizes analysed.} \label{fig:cav_runtime}
\end{figure}

Figure \ref{fig:cav_runtime} depicts the mean runtime, i.e. averaged total time per time step, (\texttt{T\textsubscript{TT}{}} from Alg. \ref{al:fs}) in log-log scale obtained with the three HPC systems for the six problem sizes. 
The runtime on the SKL and EPYC systems follows a linear decay for problem sizes N\textsubscript{e}{} $\geq$ 64 million cells suggesting $\mathtt{Hydro3D}$ scales well on these architectures. 
Conversely, when using a large number of nodes for the three smallest problem sizes, i.e. N\textsubscript{e}{} $\leq 32.8$, the average runtime plateaus as communication overheads start to increase compared to a relatively lower computing demand given the small number of grid cells per core. 
This is more noticeable in the TX2 system as, for instance, with N\textsubscript{e}{} = 8.0 the percentage of communications using two nodes is approx. 11\%, increasing to almost 90\% for the highest node count.

For most problem sizes on the TX2 cluster, there is a steady decrease in runtime with increasing number of processors although this follows a smaller reduction rate than on the other systems, indicating that the TX2 processors demonstrate less effective strong scaling in this case compared to the x86 processors. 
It is worth noting that for N\textsubscript{e}{} $\leq 8$, TX2 fails to achieve any reduction in the computation time when increasing the number of nodes beyond 8 due to communication overheads. 
For instance, for N\textsubscript{e}{} = 4.1 using 2, 8 or 25 nodes, the proportion of runtime spent on MPI communications grows from 10\%, to 48\%, and 83\% respectively, whilst for N\textsubscript{e}{} = 64 these figures are 16\%, 16\% and 45\%.
Nevertheless, for relatively large problem sizes, i.e. N\textsubscript{e}{} = 512 and 1,000, the performance and scalability of the TX2 cluster improves featuring a linear decay in runtime similar to that found on the SKL and EPYC systems.
%Note that for the largest problem sizes, simulations on a single node did not run (WHY?) and also on TX2 for $N_e = 512$.
Overall, the results of Figure \ref{fig:cav_runtime} show that the EPYC cluster provides the fastest computations for all the lid-driven cavity flow simulations.

\begin{figure}[ht!]
\centering
\includegraphics[width=0.85\linewidth]{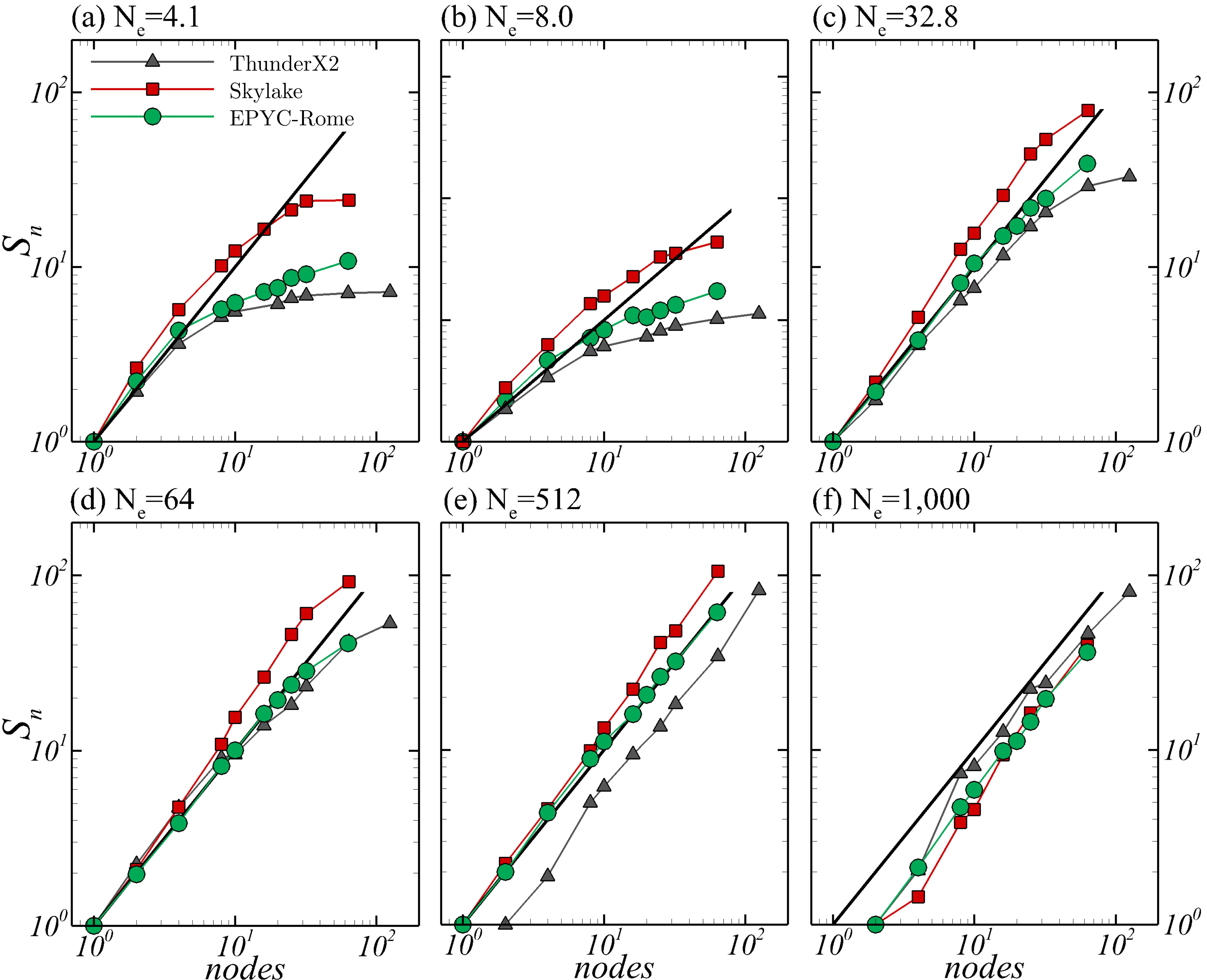}
\caption{Lid-driven Cavity flow: Results of speed up for the three HPC architectures and six grid sizes analysed. Ideal linear speedup is represented with a straight black line.} \label{fig:cav_Sn}
\end{figure} 

\begin{figure}[h!]
\centering
\includegraphics[width=0.99\linewidth]{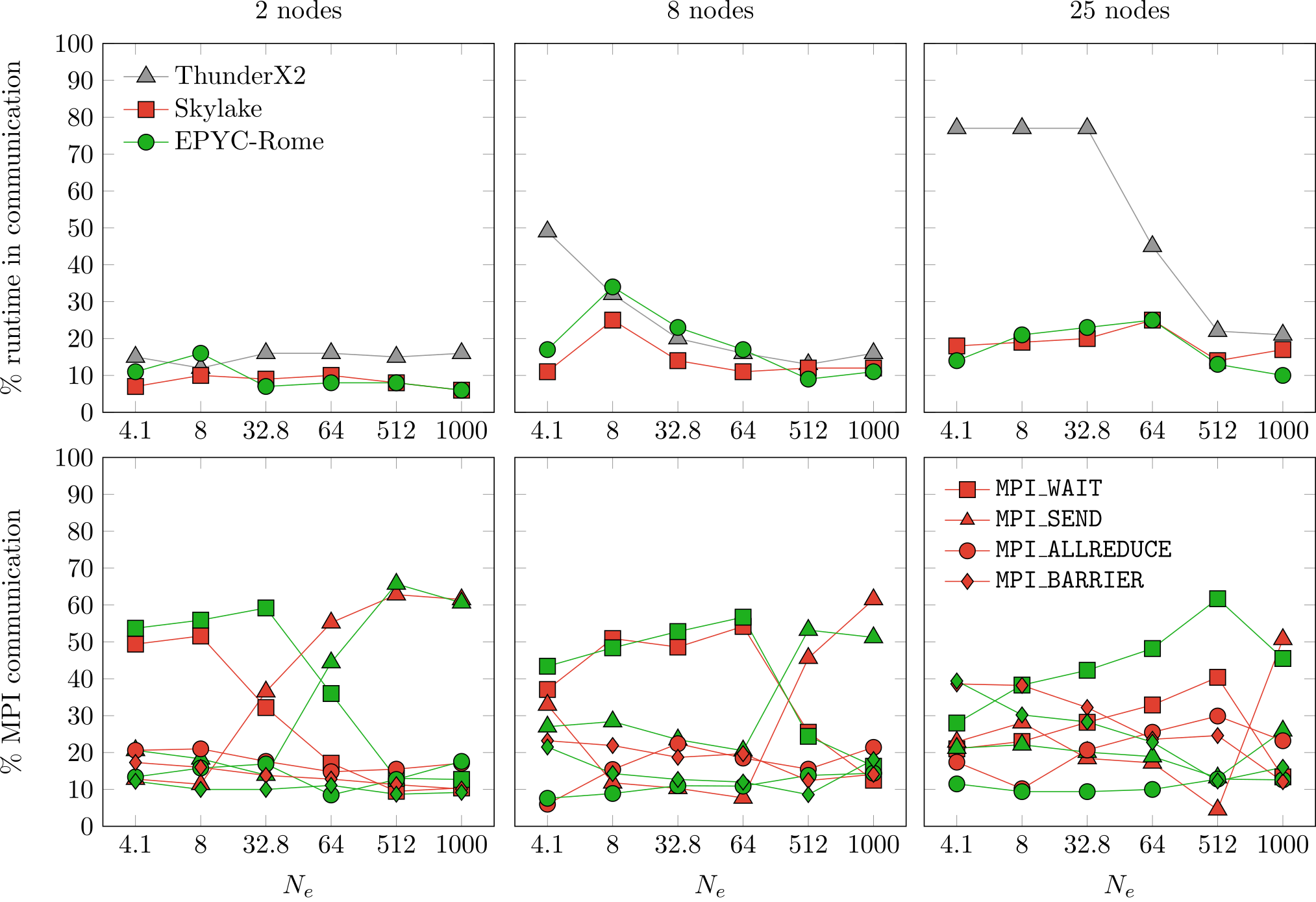}
\caption{Comparison for the lid-driven cavity flow benchmark of the percentage of runtime spent on communication (top) when running on 2, 8 and 25 nodes with the three core architectures; and distribution of MPI directives for the cases with SKL (red line) and EPYC (green line).} \label{fig:mpi_cavity}
\end{figure}

Strong scaling results are presented in Figure \ref{fig:cav_Sn} in terms of speed up $S_n = T_1 / T_n$, with $T_n$ denoting the averaged runtime per time step when using $n$ processors, in log-log scale. 
It is observed that the speed up on TX2 for small problem sizes does not increase linearly when the number of nodes is over 8, again a consequence of the large communication overhead.
Conversely, increasing N\textsubscript{e}{} improves the speed up on TX2 up to 125 nodes, achieving comparable strong scalability to that found on the SKL and EPYC systems.
For these small size problems, scalability on EPYC is inferior to that found on SKL as a result of the latter spending a lower percentage of time in communications, as shown later in Figure \ref{fig:mpi_cavity}.

%SKL processors achieve super-linearity with $S_n \geq 1.0$ in most cases, resulting from caching effects \cite{GODDEKE2013132,Yang2017}. (PABLO: CAN WE HIGHLIGHT THIS AS AN ADVANTAGE OF INTEL OVER TX2 AND EPYC-Rome?).
% UNAI: We should profile this before assuming that its the cache

Further insight into the code's strong scalability is provided in terms of parallel efficiency $E = \frac{P T_p}{Q T_q}$, with $T_p$ and $T_q$ the time per time step when using $P$ and $Q$ nodes, respectively. Ideal scalability is achieved when the parallel efficiency is 1.0, i.e. the decrease in runtime should be directly proportional to the increase in number of processors. 
However, the time spent on communication and computational processes, e.g. solving the Poisson equation with multi-grid \cite{Ouro2019CAF}, or even caching effects, could lead to non-linear $E$ growth when increasing the number of nodes.

\begin{figure}[ht!]
\centering
\includegraphics[width=0.85\linewidth]{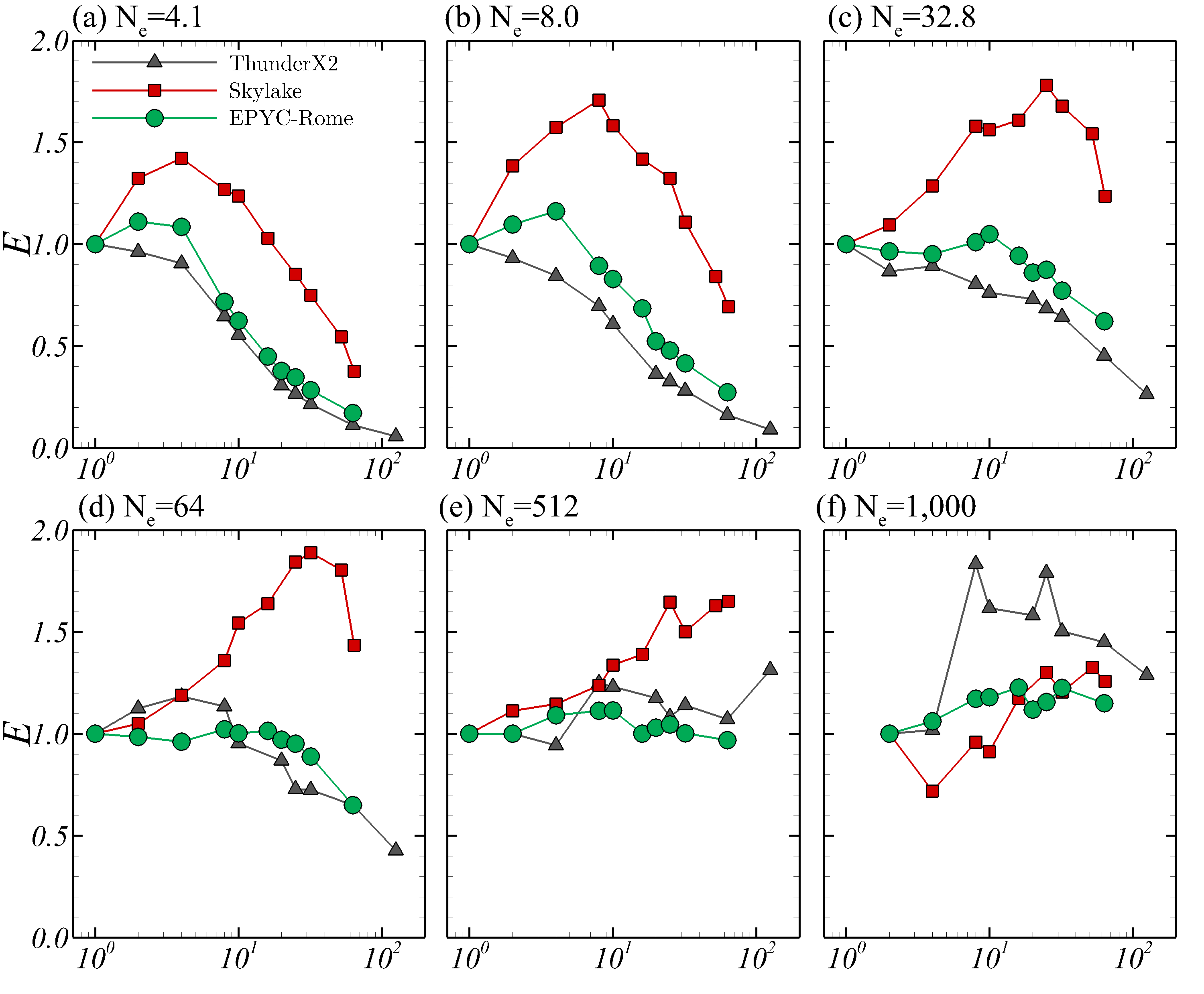}
\caption{Lid-driven Cavity flow: Results of parallel efficiency for the three HPC architectures and six grid sizes analysed.} \label{fig:cav_paral}
\end{figure} 

Figure \ref{fig:cav_paral} presents on a semi-log scale the values of $E$ for the lid-driven cavity flow simulations. There is a good parallel efficiency on SKL with increasing number of nodes for those problem sizes with N\textsubscript{e}{} $\geq$ 32.8, whilst for the two smallest problem sizes using a larger number of nodes impacts negatively on the code's scalability as $E\leq1$. 
TX2 and EPYC results show inferior parallel efficiency in problem sizes with N\textsubscript{e}{} $\geq$ 64, but exhibit improved performance when N\textsubscript{e}{} $\leq$ 64 with values of $E$ near unity.

\begin{figure}[h!]
\centering
\includegraphics[width=0.8\linewidth]{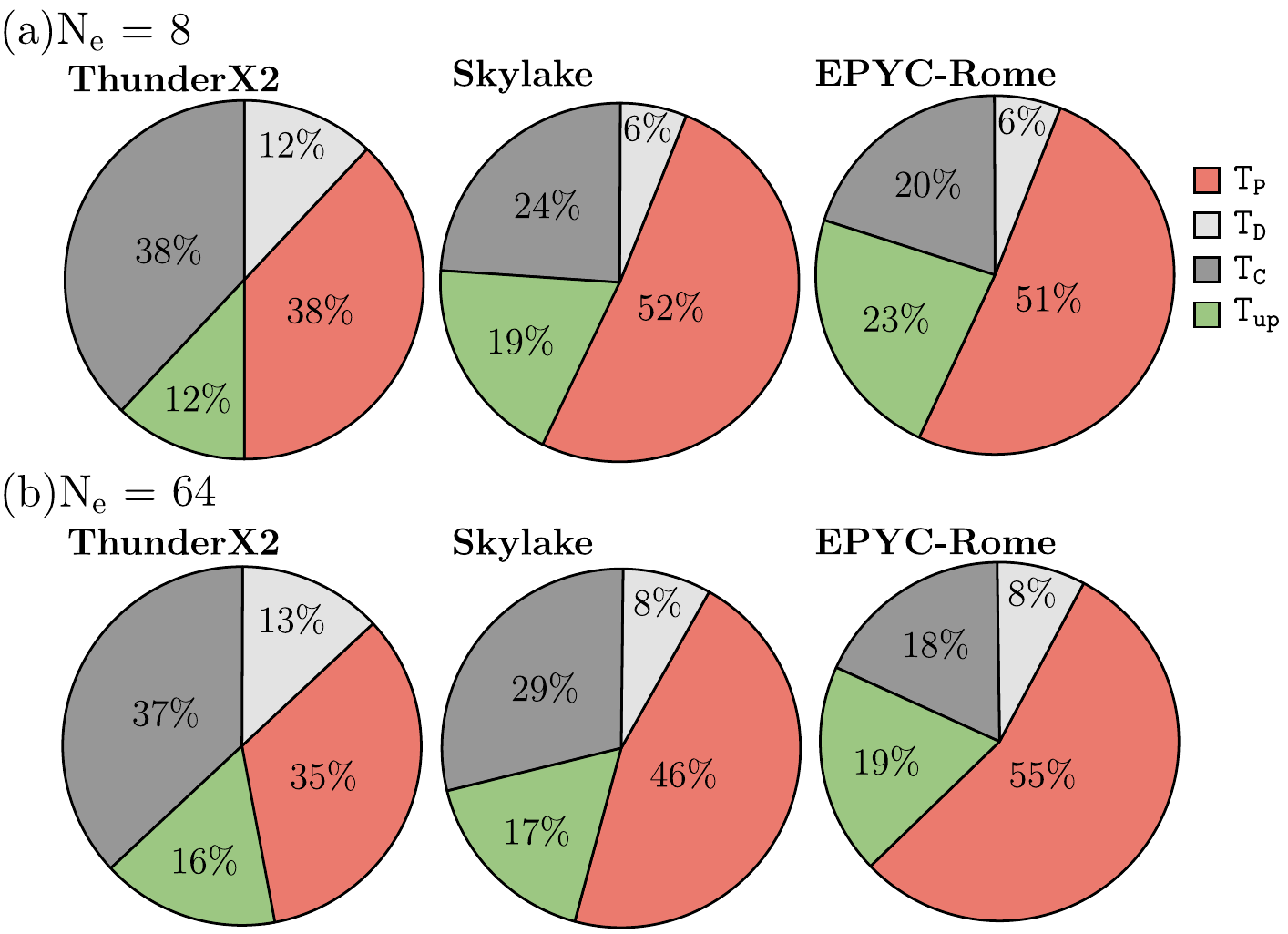}
\caption{Distribution of the main computing functions for the lid-driven cavity flow benchmark when two nodes are used. Comparison between TX2 (left), SKL (centre) and EPYC (right) for meshes N\textsubscript{e}{} = 8 (a) and 64 (b). \texttt{T\textsubscript{P}{}} = Poisson pressure solver, \texttt{T\textsubscript{D}{}} = diffusion, \texttt{T\textsubscript{C}{}} = convection, \texttt{T\textsubscript{up}{}} = velocity update.} \label{fig:pie_cavity}
\end{figure}

In order to better understand the performance results of $\texttt{Hydro3D}$, a breakdown of the compute timestep is presented in Figure \ref{fig:pie_cavity} when using two nodes and problem sizes N\textsubscript{e}{} = 8 and 64. 
In all cases, solving the iterative Poisson pressure equation (\texttt{T\textsubscript{P}{}}) takes most of the computing time, even outweighing all other subroutines combined for both SKL and EPYC systems. 
On TX2, the time spent on convection (\texttt{T\textsubscript{C}{}}) is similar to \texttt{T\textsubscript{P}{}} and larger than that obtained on the other two systems. Note that increasing the problem size has little effect on the distribution of computing times.

%\begin{table}[]
%\centering
%\begin{footnotesize}
%\begin{tabular}{l} \hline
%\end{tabular}
%\begin{tabular}{l|lll|lll|lll}
%N\textsubscript{e}{} & \multicolumn{3}{c}{2 nodes}&\multicolumn{3}{c}{8 nodes}&\multicolumn{3}{c}{25 nodes}\\ \hline
% & TX2 & SKL & EPYC & TX2 & SKL & EPYC & TX2 & SKL & EPYC \\
%4.1 & 15 & 7 & 11 & 49 & 11 & 17 & 77 & 18 & 14 \\
%8.0 & 12 & 10 &16 & 32 & 25& 34 & 77 & 19 & 21 \\
%32.8& 16 & 9 & 7 & 20 & 14 & 23 & 77& 20 & 23 \\
%64 &  16 & 10 & 8 & 16& 11& 17& 45& 25 & 25 \\
%512 & 15 & 8 & 8& 13 & 12 & 9 & 22 & 14 & 13 \\
%1,000 & 16 & 6 & 6 & 16 &12 & 11 &21&17&10 \\
%\hline
%\end{tabular}
%\end{footnotesize}
%\caption{Time spent on communications in percentage for the problem sizes and processor architectures considered.} %\label{t:cav_comm}
%\end{table}

To complement the performance breakdown, Figure \ref{fig:mpi_cavity} (top) presents the percentage of the total runtime spent in communication for the three cluster systems, for all six problem sizes using 2, 8 and 25 nodes. 
For the smaller node counts, communication accounts for less than 20\% of the simulation time in all cases, this being slightly higher on TX2. 
With 8 nodes, the N\textsubscript{e}{} = 4.1 problem size shows an increased time in communication especially for TX2, which decreases when increasing N\textsubscript{e}{} as the computing workload increases. 
However, with 25 nodes the relative time spent in communication with the x86 systems stays in the range of 20--25\% while it reaches up to 70\% for $N_e \leq$ 32.8 on TX2. 

Figure \ref{fig:mpi_cavity} (bottom) provides a more detailed communication profile from simulations on SKL and EPYC with the breakdown of the MPI time spent into synchronisation with \texttt{MPI\_WAIT} and \texttt{MPI\_BARRIER}, collective communication \texttt{MPI\_ALLREDUCE}, and point-to-point communication \texttt{MPI\_SEND}. These data were not collected for TX2 as the IPM compiler used for this analysis was not available.
%In simulations using up to 8 nodes, \texttt{MPI\_WAIT} accounts for most of the MPI time except when the problem size is large as \texttt{MPI\_SEND} increases. 
Synchronisation with \texttt{MPI\_WAIT} arises as the major communication overhead except for $N_e \geq$ 512 in which \texttt{MPI\_SEND} increases considerably as a result of communications involving many small messages distributed across a large number of cores. 
The switch from \texttt{MPI\_WAIT} to \texttt{MPI\_SEND} consuming most of the MPI communication overhead depends on the problem size and number of nodes, and also varies slightly between EPYC and SKL systems.
For this benchmark case, collective communications have a reduced impact in the overall communication time.

The results for this lid-driven cavity flow benchmark, in which 4$^{th}$-order CD is used to compute the convective fluxes, indicate that the EPYC system delivers the best computational performance whilst the SKL cluster exhibits the best strong scalability.

%%%%%%%%%%%%%%%%%%%%%%%%%%%%%%%%%%%%%%%%%%%
%\pagebreak \newpage
\subsection{Taylor-Green vortex}

For the Taylor-Green Vortex (TGV) benchmark, the computing overhead from the convective fluxes increases from using a 5$^{th}$-order WENO scheme compared to CD schemes, as the former requires the computation of optimal weights \cite{Tsoutsanis2018}. 
Analogous to the previous case, four mesh sizes are adopted with details provided in Table \ref{t:tgv_res} regarding mesh size ($\Delta {x_i}$), number of elements per spatial direction ($n_{x_i}$) and total number of grid cells in millions (N\textsubscript{e}{}). Note that this range of grid resolutions is similar to those available in the literature \cite{Fehn2018}. 

\begin{table}[h!]
 \caption{Numerical details of the meshes cases used for the TGV simulations.}
\centering
\begin{tabular}{lccccc}
\hline
 $\Delta {x_i}$ & $n_{x_i}$ & N\textsubscript{e}{} ($\cdot 10^6$) \\
\hline
0.005       & 200     & 8.0 \\
0.003125    & 320     & 32.8 \\
0.0015625   & 640     & 262.1 \\
0.00100     & 1,000     & 1,000 \\
\hline
\end{tabular} \label{t:tgv_res}
\end{table}

\begin{figure}[h!]
\centering
\includegraphics[width=0.99\linewidth]{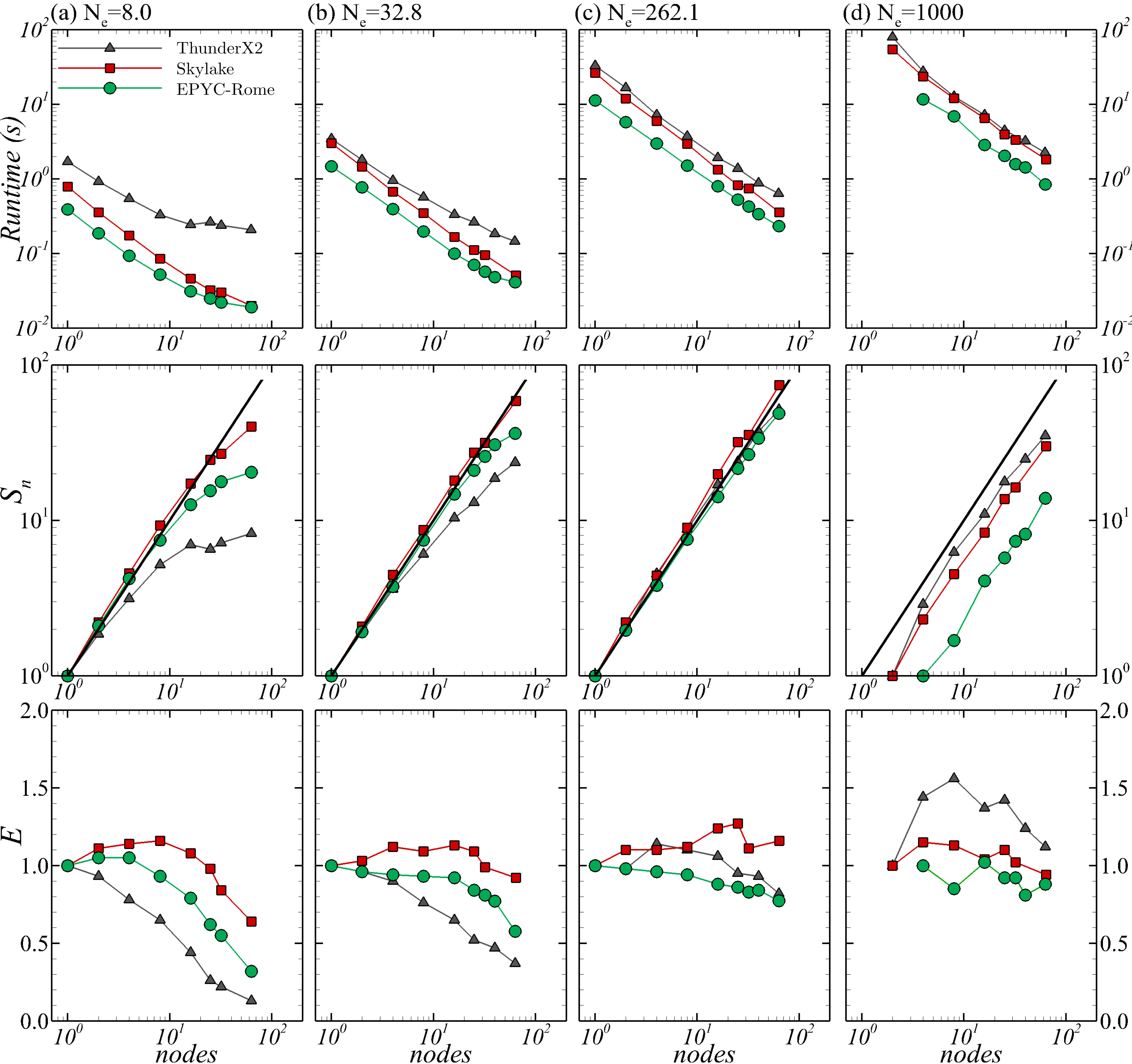}
\caption{Results of runtime (top), speed up ($S_n$) (middle), and parallel efficiency ($E$) (bottom) for the Taylor-Green vortex case.} \label{fig:tgv_res}
\end{figure}

Results of the mean runtime per time step in seconds, speed up ($S_n$) and parallel efficiency ($E$) are presented in Figure \ref{fig:tgv_res}.
For the smallest problem size, EPYC and SKL systems achieve lower runtimes than TX2 for all core counts, being up to 10$\times$ faster when 63 nodes are used. 
As with the lid-driven cavity flow benchmark, the runtime achieved with TX2 for N\textsubscript{e}{} = 8 plateaus when running on 16 or more nodes due to communication overheads.
This is also observed in the $S_n$ and $E$ plots, pointing to the need to improve the strong scalability of $\mathtt{Hydro3D}$ on TX2 for relatively small problem sizes, even though this is superior to that obtained for the lid-driven cavity flow.

With N\textsubscript{e}{} = 32.8, $\mathtt{Hydro3D}$ achieves an almost linear $S_n$ increase for each system. 
Taking a closer look at the relative performance between processors for large problem sizes, for N\textsubscript{e}{} = 262.1 SKL is on average approx. 1.4$\times$ faster than TX2 whilst EPYC is 2$\times$ and 2.6$\times$ faster than SKL and TX2, respectively. 
Based on node-to-node comparison, with N\textsubscript{e}{} = 1,000, SKL provides a 1.1$\times$ performance enhancement compared to TX2, whose improved performance in large problem sizes is due to increased memory bandwidth. 
On the other hand, EPYC is again the fastest system being 2.0$\times$ faster than SKL and 2.3$\times$ faster than TX2.
Compared to the lid-driven cavity flow case, the linearity observed in the SKL $S_n$ and $E$ curves is less apparent, resulting perhaps from the WENO weights becoming the most expensive subroutine over the Poisson pressure solver.
%, as depicted from Figure \ref{fig:pie_tgv}. 

\begin{figure}[h!]
\centering
\includegraphics[width=0.8\linewidth]{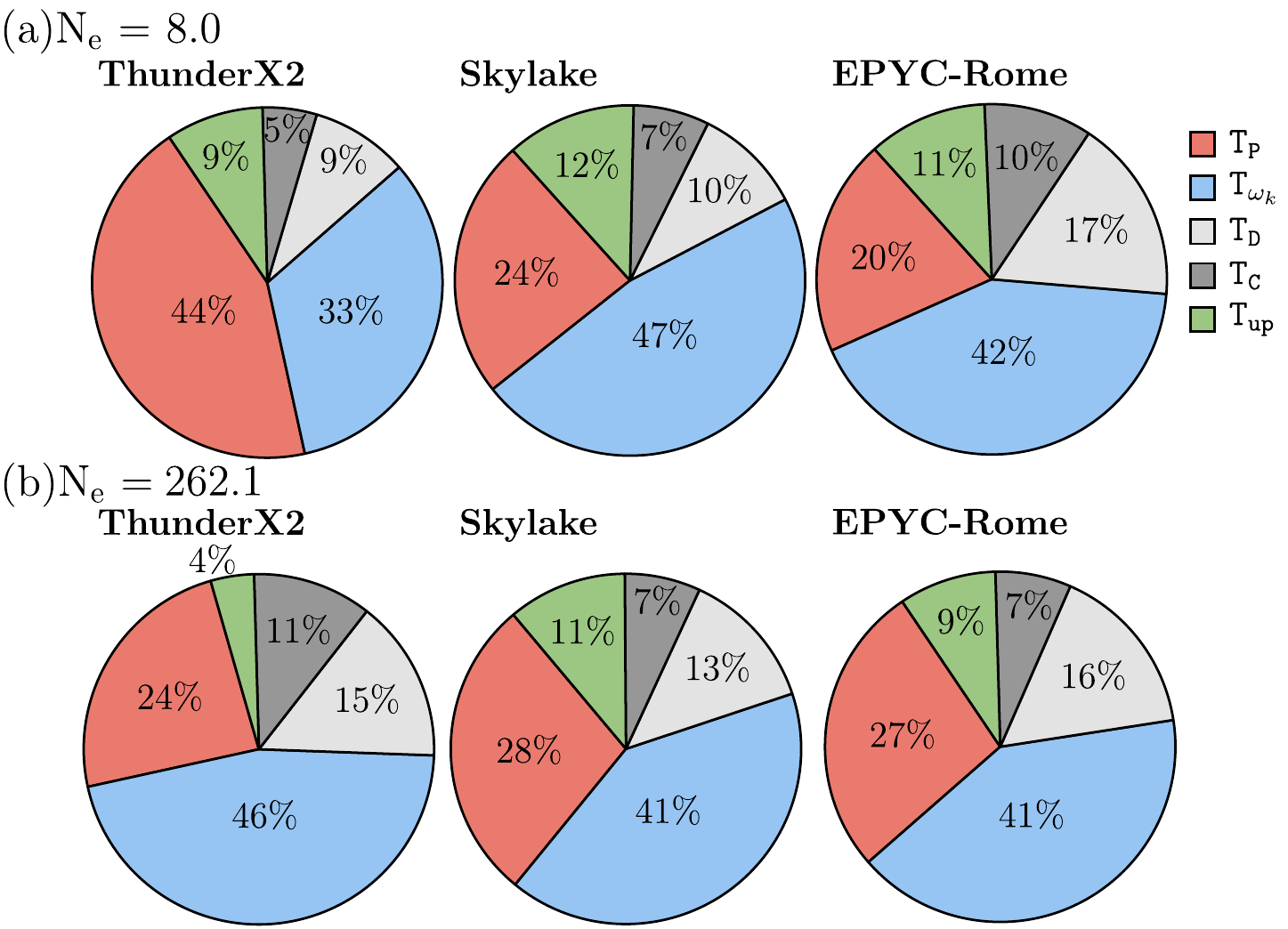}
\caption{Breakdown of the main computing functions for the TGV benchmark when two nodes are used. Comparison between TX2 (left), SKL (centre) and EPYC (right) for meshes N\textsubscript{e}{} = 8.0 (a) and 262.1 (b).} \label{fig:pie_tgv}
\end{figure}

Figure \ref{fig:pie_tgv} shows the breakdown of the runtime spent on the main computing subroutines, namely Poisson pressure solver (\texttt{T\textsubscript{p}{}}), diffusion (\texttt{T\textsubscript{D}{}}), convection (\texttt{T\textsubscript{c}{}}), velocity update (\texttt{T\textsubscript{up}{}}) and WENO weights (\texttt{T\textsubscript{$\omega_k$}{}}), for N\textsubscript{e}{} = 8 and 262.1 and using two nodes. 
The computing workload is similarly distributed across systems with the multi-grid pressure solver and WENO weights taking approx. 18--20\% and 41--47\% of the computing time, except on TX2 with N\textsubscript{e}{} = 8. 
This is attributed to the lower computing capability of TX2, as seen in Section \ref{S:6.0}, increasing the time required for iteratively solving the pressure equation.
Although not included here for sake of brevity, results for the other two meshes and larger node count present a similar distribution, pointing to the codes' subroutines scaling well. 

\begin{figure}[h!]
\centering
\includegraphics[width=0.99\linewidth]{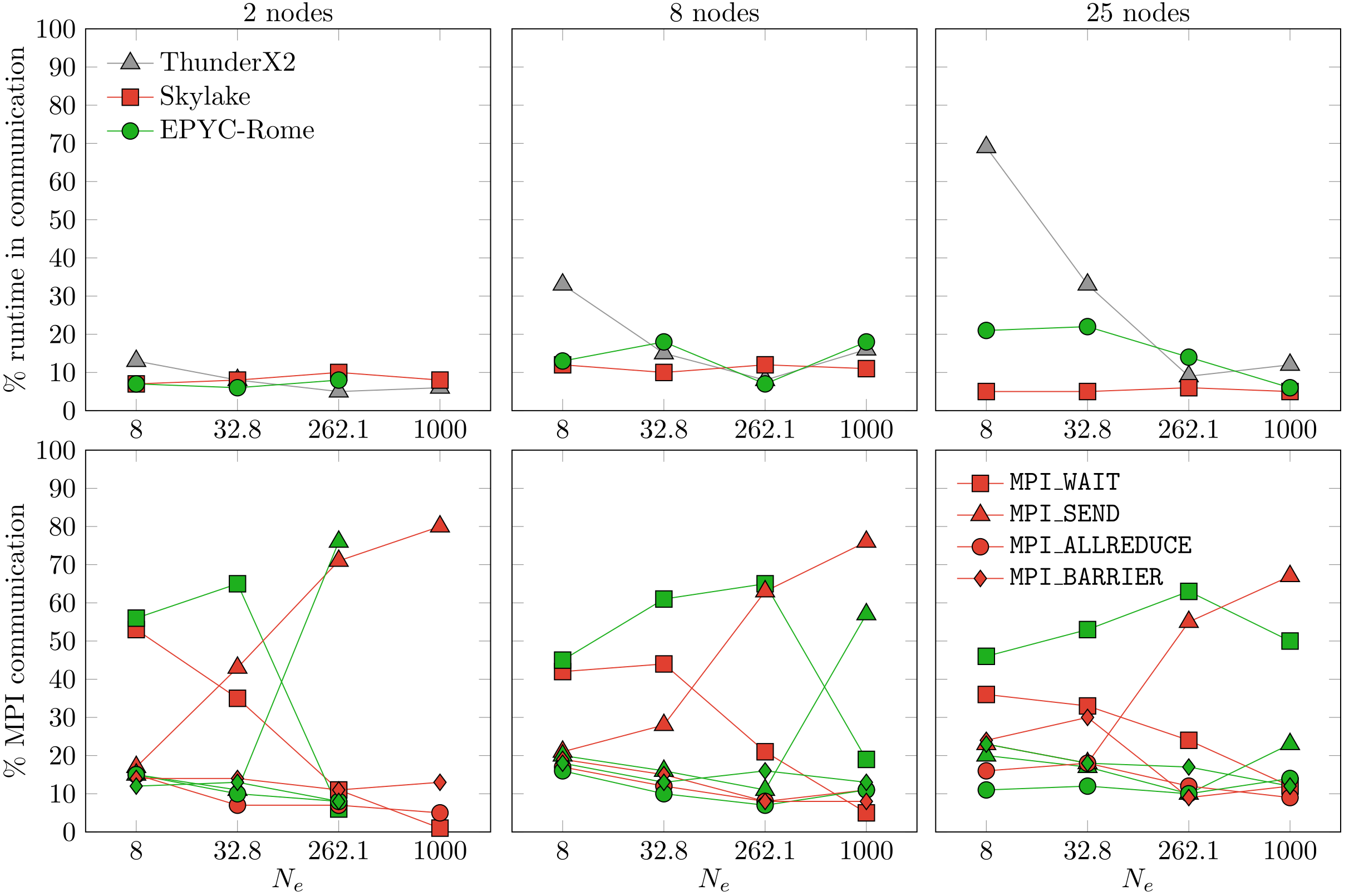}
\caption{Comparison of the percentage of runtime spent on communication (top) when running on 2, 8 and 25 nodes for the TGV benchmark; and distribution of MPI directives for the cases with Skylake (red line) and EPYC-Rome (green line).} \label{fig:mpi_tgv}
\end{figure}

The communication profiles are presented in Figure \ref{fig:mpi_tgv}, showing that for 2 nodes the simulations spend little time in communications. 
For 8 and 25 nodes, the relative weight of data exchange increases, especially for small problem sizes on TX2, with the percentage of communication notably increasing. 
In relation to the MPI directives, \texttt{MPI\_WAIT} and \texttt{MPI\_WAIT} consume most of the communication time with a balance similar to that observed in the lid-driven cavity flow results when increasing the number of nodes.

%%%%%%%%%%%%%%%%%%%%%%%%%%%%%%%%%%%%%%%%%%%
%\pagebreak \newpage
\subsection{Solitary wave simulation with level-set method}

The third benchmark presents the solitary wave simulation using LSM.
The previous TGV benchmark showcased the computing overhead from the 5$^{th}$-order WENO scheme, which is now expected to increase given the computation of both convective fluxes and LSM advection and reinitialisation equations.
In this benchmark, the travelling wave is simulated using three grid resolutions with the details of mesh resolution ($\Delta {x_i}$) that is uniform across the domain, the number of divisions per spatial directions ($n_{x_i}$) and the total number of grid cells (N\textsubscript{e}{}, in millions) presented in Table \ref{t:wave_res}. 

\begin{table}[h!]
 \caption{Numerical details of the meshes used for the solitary wave benchmark.} 
\centering
\begin{tabular}{cccccc}
\hline
 $\Delta {x_i}$ & $n_{x}$ & $n_{y}$ & $n_{z}$ & N\textsubscript{e}{} ($\cdot 10^6$) \\
\hline
0.0100   & 1,280 & 40 & 30      & ~1.5 \\
0.0050   & 2,560 & 80 & 60      & 12.3 \\
0.0025   & 5,120 & 160 & 120    & 98.3 \\
\hline
\end{tabular}\label{t:wave_res}
\end{table}

\begin{figure}[h!]
\centering
\includegraphics[width=0.95\linewidth]{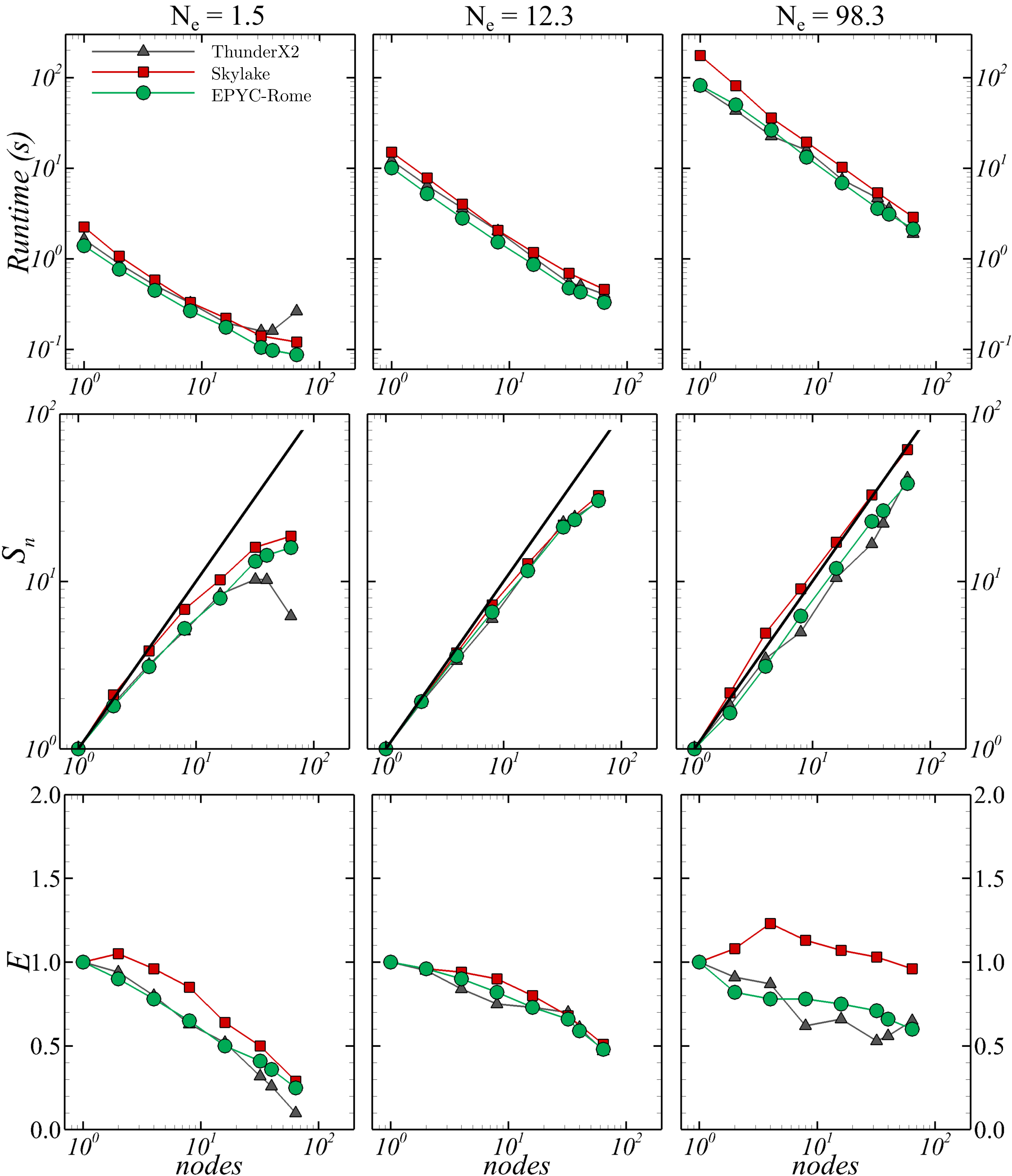} %.png}
\caption{Results of averaged runtime per time step values in seconds (top), speed up (mid) and parallel efficiency (bottom) for the three HPC architectures and three grid sizes analysed in the solitary wave benchmark.} \label{fig:wave_runtime}
\end{figure}

The mean time per time step obtained for the three problem sizes on each HPC system is shown in Figure \ref{fig:wave_runtime} (top) using a log-log scale. 
All systems show an almost linear decrease in runtime values with increasing node count. 
Unlike the former benchmarks, the percentage of communication in the solitary wave simulation is less than 20\% of the total runtime for configurations running on the maximum number of nodes.
$\mathtt{Hydro3D}$ achieves its best performance on the EPYC cluster for all cases, although it is noticeable that simulation times obtained with TX2 outperform those from SKL by approx. 1.4--1.6$\times$, especially in the highest resolution case, in which TX2 and EPYC systems achieve similar runtimes. 
In this benchmark, both systems spend a relatively lower percentage of time in communications than does SKL, thereby enhancing inter-node communications and proving advantageous when using LSM as \texttt{MPI\_ALLREDUCE} is called several times triggering the time in data exchange.

\begin{figure}[h!]
\centering
\includegraphics[width=0.8\linewidth]{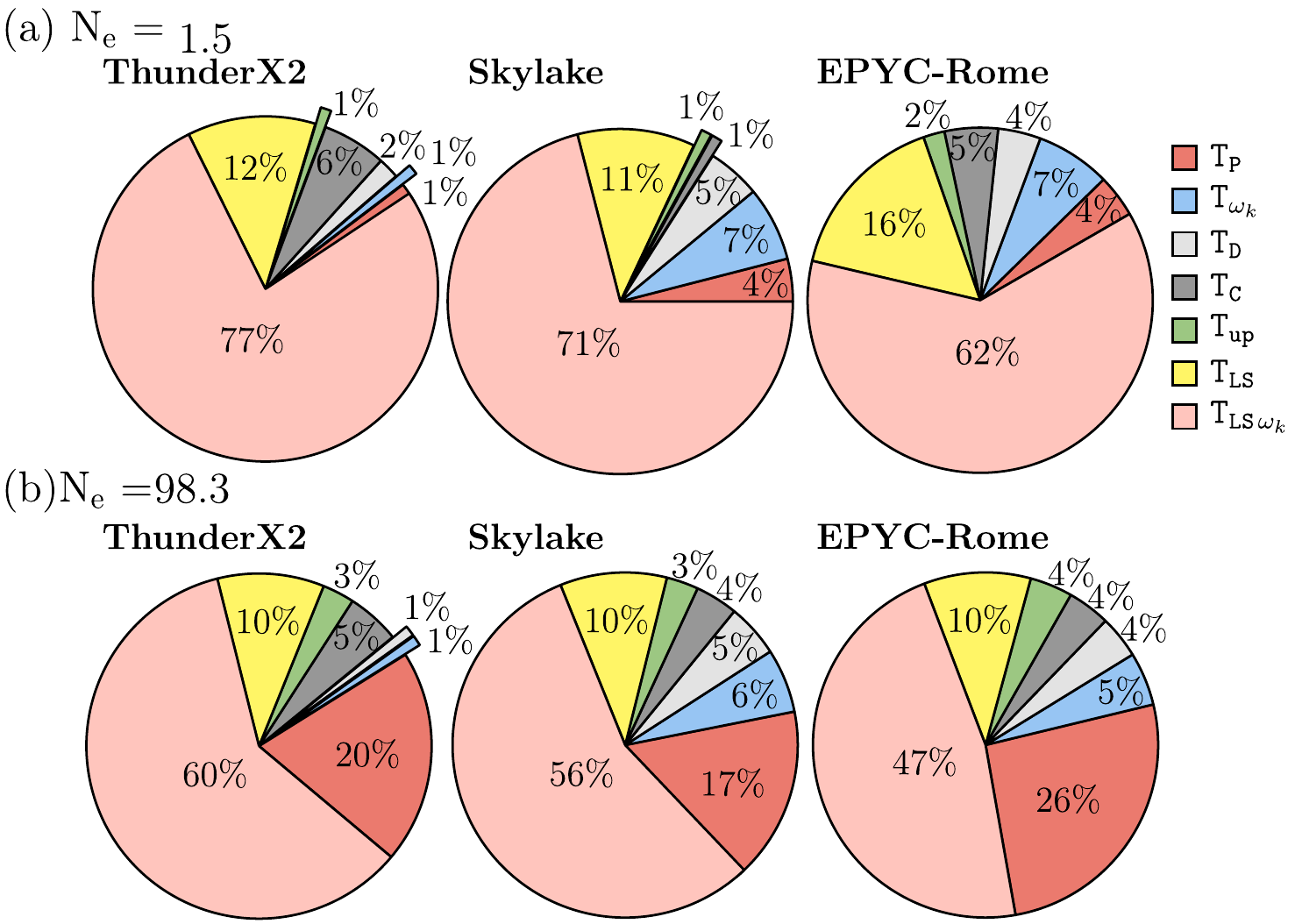} %.png}
\caption{Breakdown of the main computing components for the solitary wave benchmark case when running on two nodes. Comparison between TX2 (left), SKL (centre) and EPYC (right) for meshes N\textsubscript{e}{} = 1.5 (a) and 98.3 (b). \texttt{T\textsubscript{P}{}} = Poisson pressure solver, \texttt{T\textsubscript{$\omega_k$}{}} = WENO weights, \texttt{T\textsubscript{D}{}} = diffusion, \texttt{T\textsubscript{C}{}} = convection, \texttt{T\textsubscript{up}{}} = velocity update, \texttt{T\textsubscript{LS}{}} = LSM, and \texttt{T\textsubscript{LS\_$\omega_k$}{}} WENO weights in LSM.}  \label{fig:pie_wave}
\end{figure}

The resulting speed up ($S_n$) and parallel efficiency ($E$) for the solitary wave case are shown in Figure \ref{fig:wave_runtime} (mid and bottom). 
For N\textsubscript{e}{} = 1.5, $\mathtt{Hydro3D}$ features less than optimal strong scalability with $S_n$ being below the ideal linear performance increase and values of $E$ steadily decreasing with increasing node count for all systems. 
The communication-to-computation ratio grows almost linearly from 7\% to 19\% when increasing from 1 to 64 nodes irrespective of the problem size, indicating that communications are not the major overhead.
Instead, resolving the LSM with the present grid resolutions requires 15 iterations in the $\phi$ re-initialisation which triggers the overhead from computing WENO reconstructions \cite{Tsoutsanis2018}, as observed in Figure \ref{fig:pie_wave}.

Increasing the resolution up to N\textsubscript{e}{} = 12.3 provides an improvement in scalability, as shown by the speed up curve, and results for N\textsubscript{e}{} = 98.3 show that SKL follows a linear increase in $S_n$, attaining values of $E$ above one.
TX2 and EPYC feature an almost linear $S_n$ increase below the ideal threshold and steadily decrease in parallel efficiency with increasing node count. 
Overall, these results suggest that \texttt{Hydro3D} demonstrates good scalability depending on the overhead from the LSM re-initialisation that appears as the most computationally demanding task in the solitary wave benchmark.

The breakdown of the computing time (again excluding communications) is presented in Figure \ref{fig:pie_wave} for the smaller and larger problem sizes using two nodes, including time spent on the LSM advection equations (\texttt{T\textsubscript{LS}{}}) with the time spent on computing the WENO weights for $\phi$ accounted separately as \texttt{T\textsubscript{LS\_$\omega_k$}{}} (see Alg. \ref{al:lsm}).
For the small problem size, Figure \ref{fig:pie_wave}(a), \texttt{T\textsubscript{LS\_$\omega_k$}{}} represents between 62--77\% of the total computing time, with the rest of the LSM workload, i.e. \texttt{T\textsubscript{LS}{}}, in the range of 11--16\%.
For N\textsubscript{e}{} = 98.3, the relative contribution from the pressure solver (\texttt{T\textsubscript{P}{}}) increases to 17--26\% of the total cost, in turn decreasing the percentage attributed to computing LSM.
These results outline the high additional computational cost of LSM for multi-phase flows, which normally requires very fine grid resolutions and time steps to guarantee numerical stability.
Considering the workload distribution, the code spends a similar proportion of time in the various subroutines when running on the SKL and EPYC systems, whilst on TX2 the overall time spent in convection, diffusion, velocity update and WENO weights for the convection scheme are much reduced. 

%\textbf{due its lower clock frequency that leads the code to spend more time resolving the LSM and pressure} (PABLO: DOES THIS MAKE SENSE?).
% UNAI: Might need a profile to verify this.

\begin{figure}[h!]
\centering
\includegraphics[width=0.99\linewidth]{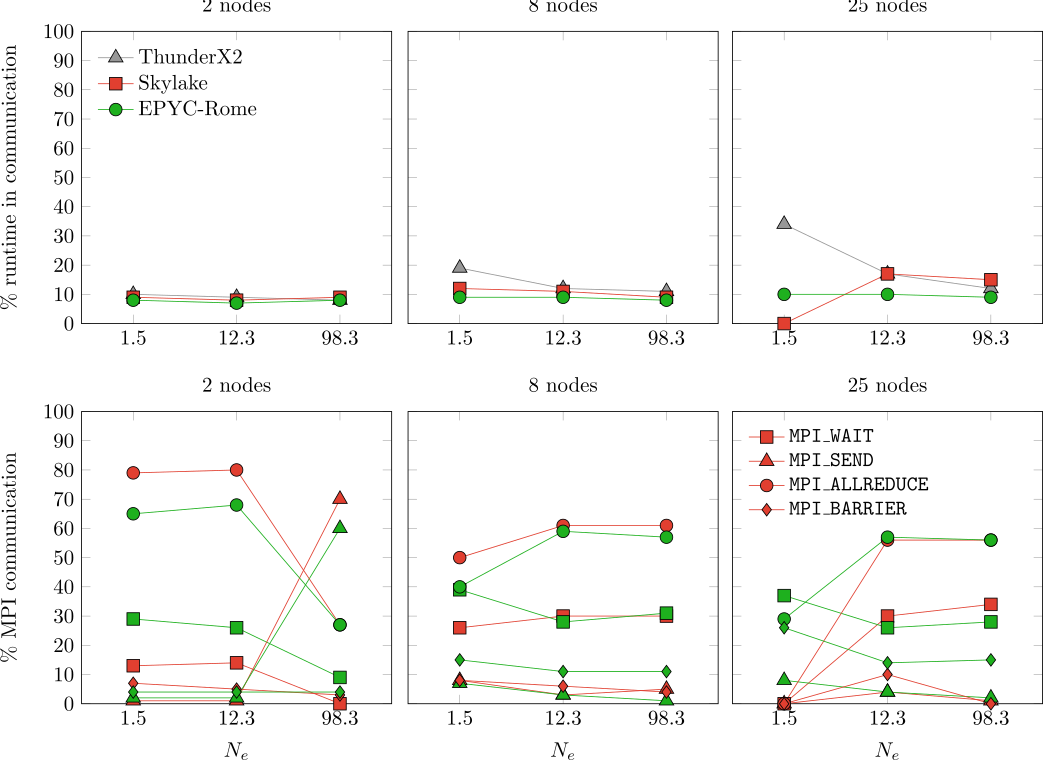}
\caption{Comparison of the percentage of runtime spent on communication (top) when running on 2, 8 and 25 nodes of the three systems; and distribution of MPI functions for the cases with SKL (red line) and EPYC (green line)systems. Solitary Wave benchmark case.} \label{fig:mpi_wave}
\end{figure}

Profiles of communication overhead over the total runtime and MPI functions for this benchmark are presented in Figure \ref{fig:mpi_wave} for simulations with 2, 8 and 25 nodes and the three N\textsubscript{e}{} adopted.
The overall contribution of the communication operations in the simulation is small for 2 and 8 nodes and slightly higher for 25 nodes.
Only for N\textsubscript{e}{} = 1.5 does the code on TX2 experience an increase in communication overheads analogous to the results obtained for the other benchmark cases. 
The MPI directive consuming most of the communication time is the collective \texttt{MPI\_ALLREDUCE} required during the LSM, as indicated in Alg. \ref{al:lsm}. 
The synchronisation primitive \texttt{MPI\_WAIT} is the second MPI function consuming most communication time, which represents approx 2--6\% of the total runtime for each case i.e., $\mathtt{Hydro3D}$ spends very little time constrained by communication in this benchmark.

\section{Conclusions} \label{S:conclusions}

This paper compares the performance of three cluster systems characterised by distinctive node architectures, namely Intel Skylake (SKL), ARMv8.1 ThunderX2 (TX2) and AMD EPYC-Rome (EPYC), in applications based on a state-of-the-art, in-house Computational Fluid Dynamics (CFD) code, $\mathtt{Hydro3D}$.
This study investigated the code's multi-node performance and scalability in three benchmarks comprising several problem sizes and using up to 8,000 cores.
The benchmark cases considered were the lid-driven cavity flow, the Taylor-Green vortex and the propagation of a solitary wave in a numerical tank, designed to perform computations of increasing difficulty and expense.
The computing workload is varied changing the computation of the convective fluxes with 4$^{th}$-order central differences and 5$^{th}$-order WENO schemes for the first two cases respectively, whilst in the wave simulation the level-set method equations are computed, also with the 5$^{th}$-order WENO scheme.

The EPYC-based system provides the fastest computing times for all benchmark cases irrespective of the problem size, attributed to the 64 cores per node in comparison to the 40-core SKL nodes and larger clock frequency than the TX2 processors. 
TX2 performance for small problem sizes features sub-optimal strong scalability as the communications overhead notably increases. 
However, increasing the number of grid cells leads to reduced differences between TX2 and SKL with both demonstrating similar runtime values, although these were greater than those obtained with EPYC processors.
For the solitary wave case, TX2 achieves better computational performance than SKL processors for all problem sizes. 
Nonetheless, a notable feature in most cases is that $\mathtt{Hydro3D}$ needs to improve its strong scalability on the TX2, especially when the number of grid cells is relatively small, resulting from an overhead in blocking MPI communications.

This study suggests that the major drawback when adopting high-order WENO schemes is their larger computing overhead compared to central differences. Results show that in simulations with the WENO scheme the code's performance on TX2 is similar to that on SKL processors. 
This suggests that CFD codes using finite volumes or finite elements in unstructured grids relying on mesh connectivity can demonstrate good performance when running on TX2 processors, as such schemes are computationally more demanding than central differences.
%due to the flux computation at the cell faces as well as unstructured meshes listing operations.

Results shown here are of interest to the HPC and CFD communities, both for those working on direct-numerical and large-eddy simulation applications whose computational cost is extremely high, as well as those doing engineering-oriented simulations. 
The present study is based on the performance of a structured-grid CFD code, using central differences as widely used numerical schemes to compute the viscous and convective fluxes in LES and DNS. 

Considering the maturity of the ARM ecosystem for HPC, we note that $\mathtt{Hydro3D}$ required no modification of the codebase to either compile or run on the TX2 system. The Cray compilers and other performance tools used in these tests e.g., profilers and debuggers, show a production-level maturity. From the user perspective, they offer a similar interface to other commercial HPC tools and produced no issue or error during their use.

Future work will develop in two directions: first, a study of the energy usage of these systems. The present study has centred on raw computing times, but energy efficiency is of increasing concern in the field of HPC and this will be explored with the balance between FLOP/s and Watts. 
Second, an increased focus on single-node performance. The current work has concentrated on the scalability and bottlenecks at the cluster/multi-node level and the next step will be to explore in greater detail the performance of the code at the node level.

%Overall, this paper demonstrates that EPYC-Rome processors deliver excellent code performance for CFD applications, along with the growing maturity of the emerging ARM ThunderX2 systems, on the verge of directly competing with Intel for real engineering applications on an in-house CFD code.
%(PABLO: WE NEED MORE REFERENCES TO THE DIFFERENT CHARACTERISTICS BETWEEN CHIPS IN TERMS OF MEMORY BANDWIDTH, NUMBER OF MEMORY CHANNELS, ETC.)

\section*{Acknowledgements} 
The authors would like to acknowledge the support of the Supercomputing Wales project, which is partially funded by the European Regional Development Fund (ERDF) via the Welsh Government. The Isambard project is funded by UK's EPSRC (EP/P020224/1), the GW4 alliance, the Met Office, Cray and Arm. This work was partially funded by the GW4 Data Science research project "Evaluating the performance of a large-eddy simulation code in 64-bit ARM processors". 
The authors acknowledge Prof. Simon McIntosh-Smith for his support and Drs. Ade Fewings, Thomas Green and James Price for helping in the setup and debugging of the code.

\bibliographystyle{model1-num-names}
\bibliography{OuLPGU_Original.bib}
\end{document}